%% file: HIG-12-033_temp.tex
\pdfoutput=1

\documentclass[11pt,twoside,a4paper,cmspaper,final,collab]{cms-tdr}
\def\svnVersion{185180}\def\cmsCernNo{2012-375}\def\cmsCernDate{\today}\def\cmsMessage{Submitted to Physics Letters B}

\begin{document}\cmsNoteHeader{HIG-12-033}

\hyphenation{had-ron-i-za-tion}
\hyphenation{cal-or-i-me-ter}
\hyphenation{de-vices}

\RCS$Revision: 178753 $
\RCS$HeadURL: svn+ssh://svn.cern.ch/reps/tdr2/papers/HIG-12-033/trunk/HIG-12-033.tex $
\RCS$Id: HIG-12-033.tex 178753 2013-03-29 14:13:52Z alverson $
\newlength\cmsFigWidthHalf\setlength\cmsFigWidthHalf{0.48\textwidth}
\newlength\cmsFigWidth
\ifthenelse{\boolean{cms@external}}{\setlength\cmsFigWidth{0.48\textwidth}}{\setlength\cmsFigWidth{0.75\columnwidth}}
\ifthenelse{\boolean{cms@external}}{\providecommand{\cmsLeft}{top}}{\providecommand{\cmsLeft}{left}}
\ifthenelse{\boolean{cms@external}}{\providecommand{\cmsRight}{bottom}}{\providecommand{\cmsRight}{right}}
\newcommand{\CLs}{\ensuremath{\mathrm{CL_s}\xspace}}
\newcommand{\mA}{\ensuremath{M_{\textrm{A}}}\xspace}
\newcommand{\mh}{\ensuremath{M_{\textrm{h}}}\xspace}
\newcommand{\mH}{\ensuremath{M_{\textrm{H}}}\xspace}
\newcommand{\mPhi}{\ensuremath{M_{\phi}}\xspace}
\newcommand{\mhmax}{\ensuremath{m_{\textrm{h}}^{\text{max}}}\xspace}
\cmsNoteHeader{HIG-12-033} % This is over-written in the CMS environment: useful as preprint no. for export versions
\title{Search for a Higgs boson decaying into a b-quark pair and produced in association with b
    quarks in proton-proton collisions at 7\TeV}

\author{The CMS Collaboration}

\date{\today}

\abstract{
A search for a neutral Higgs boson decaying to a pair of b quarks, and
     produced in association with at least one additional b quark, is presented.
     Multijet final states with three jets identified as originating from b quarks, at least one of which may include a non-isolated muon, are studied.
     The data used in this analysis correspond to an integrated luminosity of 2.7--4.8\fbinv, collected by the CMS experiment in proton-proton
     collisions at the LHC at a center-of-mass energy of 7\TeV. This search is particularly
     sensitive to Higgs bosons in scenarios of the Minimal Supersymmetric Model (MSSM)  with large values of $\tan \beta$.
     No excess over the predicted background from standard model processes is observed.
     Stringent upper limits on cross section times branching fraction
     are derived and interpreted as bounds in the MSSM
     $\tan\beta$ and \mA parameter-space.
     Observed 95\% confidence level upper limits reach as low as $\tan\beta\approx18$ for $M_\textrm{A}\approx 100\GeV$.
}

\hypersetup{%
pdfauthor={CMS Collaboration},%
pdftitle={Search for a Higgs boson decaying into a b-quark pair and produced in association with b quarks in proton-proton collisions at 7 TeV},
pdfsubject={CMS},%
pdfkeywords={CMS, physics, Higgs, MSSM}}

\maketitle %maketitle comes after all the front information has been supplied

\section{Introduction}\label{sec:Intro}

The electroweak symmetry breaking mechanism of the standard model (SM)
predicts the existence of a neutral scalar boson, the Higgs
particle.
A boson has been recently discovered, with a mass around
125\GeV~\cite{Aad:2012gk,Chatrchyan:2012gu} and properties consistent with those expected for the SM Higgs boson. However, its exact properties
and the detailed structure of the Higgs sector still need further investigation. Moreover, the mass of the Higgs boson is quadratically divergent at high energies~\cite{Witten-hierarchy}. Supersymmetry~\cite{Martin97} is a well known extension to the SM which allows the cancellation of this divergence.

In contrast to the SM, the Minimal Supersymmetric Standard Model
(MSSM)~\cite{ref:SUSY} features two scalar Higgs doublets,
giving rise to three neutral Higgs bosons, collectively denoted as $\phi$, and
two charged ones, $\PH^{\pm}$. Two of the neutral bosons are CP-even (h, H) and
one is CP-odd (A).
In this context, the recently discovered boson with a mass near 125\GeV might be interpreted as
one of the neutral CP-even states.
At tree level, two parameters, conventionally
chosen as the mass of the pseudoscalar Higgs boson $\mA$ and the ratio of
the vacuum expectation values of the two Higgs doublets, $\tan \beta = v_2/v_1$, define the Higgs sector in the MSSM.
For $\tan \beta$ larger than unity, the Higgs field couplings to
up-type particles are suppressed relative to the SM, while the couplings to down-type particles
are enhanced by a factor of $\tan \beta$. In addition, the mass
$\mA$ is expected to be nearly degenerate with either $\mh$ or $\mH$.
Therefore, the combined cross section of
Higgs boson production in association with b quarks is effectively enhanced by a factor
${\approx}2\tan^2\beta$. Moreover, the decay into b quarks has a very high branching fraction
(${\approx}90\%$), even at large values of the Higgs boson mass \mA. The sensitivity for a
SM Higgs boson search for the corresponding channel is negligible given the small cross section.

Recent results at the Large Hadron Collider (LHC) on the  $\phi\to\tau\tau$  decay mode~\cite{ref:Chatrchyan201268,Aad:2012yfa} provide
stringent constraints on $\tan\beta$, complementing previous results from the LEP experiments~\cite{Schael:2006cr} and superseding those from the Tevatron experiments~\cite{Aaltonen:2009vf,Abazov:2008hu,Abazov2012569}.
Similar searches in the $\phi\to\bbbar$ decay mode have also been performed by the
CDF and D0 experiments~\cite{PhysRevD.86.091101} at the Tevatron
collider.
An excess of events of ${\approx}2$ standard deviations with respect to
the expectations from SM background have been reported by both experiments for a resonance in the  mass range 100--150\GeV.

In this Letter we present a search for MSSM neutral Higgs bosons produced
in association with at least one b quark, and decaying into a pair of b
quarks.
Prospects for this channel at the LHC have been studied in Refs.~\cite{Ball:2007zza,Carena:2011a}.
This analysis is performed using 2.7--4.8\fbinv of proton-proton collisions with a center-of-mass energy of 7\TeV
collected in 2011 by the Compact Muon Solenoid (CMS) detector at the LHC.
The dominant background is the production of heavy-flavor multijet
events containing either three b jets, or two b jets plus a third jet originating from either a charm or
a light-flavor parton, which is misidentified as a b jet.

A signal is searched for in final states characterized either purely
by jets (``all-hadronic'') or with an additional non-isolated muon
(``semileptonic''). Events are selected by specialized
triggers that include online algorithms for the identification of b
jets to tackle the large multijet production rate at the LHC.
The common analysis strategy is to search, in events identified as having at least three b jets, for a peak in the invariant mass distribution of the two leading
b jets, \ie those having the largest transverse momentum, over the large multijet background.
A key point of both analyses
is the estimation of the background using control data samples, which is addressed with
different methods. The two analyses reach similar
sensitivity to the MSSM Higgs scenarios described. The corresponding data sets are largely exclusive, and the small overlap is
removed for the combined results.

\section{The CMS experiment}\label{sec:Apparatus}

The central feature of the CMS detector is a
superconducting solenoid of 6\unit{m} internal diameter, providing a magnetic field of
3.8\unit{T}.
Within the field volume, the inner tracker is formed by a silicon pixel and
strip tracker.  It measures charged particles within the pseudorapidity range
$\abs{\eta}< 2.5$.
The pseudorapidity is defined as $\eta=-\ln(\tan(\theta/2))$ and $\theta$ is the polar angle, while $\phi$ is the azimuthal angle in radians.
The tracker provides an impact parameter resolution of approximately $15\mum$
and a resolution on transverse momentum (\pt) of about 1.5\% for 100\GeV
particles. Also inside the field volume are a crystal electromagnetic
calorimeter and a brass/scintillator hadron calorimeter. Muons are
measured in gas-ionization detectors embedded in the iron flux return yoke, in the pseudorapidity range
$\abs{\eta}< 2.4$, with detector planes made using three technologies: drift
tubes, cathode-strip chambers, and resistive-plate chambers. Matching muons to
tracks measured in the silicon tracker results in a transverse momentum
resolution between 1\% and 5\%, for \pt values up to 1\TeV.
Extensive forward calorimetry complements the coverage provided by the barrel
and endcap detectors.
A more detailed description of the CMS detector can be found in Ref.~\cite{Chatrchyan:2008zzk}.

\section{Event reconstruction and simulation}\label{sec:EventDesc}

The CMS particle-flow event
reconstruction~\cite{CMS-PAS-PFT-09-001,CMS-PAS-PFT-10-001} is used
for optimized reconstruction and identification of all particles in
the event,
\ie electrons, muons, photons, charged hadrons, and neutral hadrons, with an extensive combination of all CMS detectors systems.

The reconstructed primary vertex with the largest $\pt^2$-sum of its associated tracks is selected
and used as reference for the other physics objects.

Jets are reconstructed using the anti-\kt algorithm
\cite{Cacciari:2008gp} from particle-flow objects
with a radius parameter $R = 0.5$ in the rapidity-azimuthal angle space.
Each jet is required to have more than one track
associated to it, and to have electromagnetic and hadronic energy
fractions of at least 1\% of the total jet energy.
Additional proton-proton interactions within the same bunch crossing (pileup) affect the
jet momentum reconstruction.  To mitigate this effect, a
track-based algorithm that removes all charged hadrons not
originating from the primary interaction is used. In addition,
a calorimeter-based algorithm evaluates the energy density in the
calorimeter from interactions not related to the primary vertex, and
subtracts it from the reconstructed jets in the event. Additional
jet energy corrections~\cite{Chatrchyan:2011ds} are applied.

Muons are reconstructed using both the inner silicon tracker and the outer muon
system~\cite{1748-0221-7-10-P10002}, and by performing a global track fit seeded by
signals in the muon system.

The combined secondary vertex (CSV) algorithm \cite{CMS-PAS-BTV-12-001} is used in the offline
identification of b jets. The CSV algorithm uses information on track impact
parameter and secondary vertices in a jet combined in a
likelihood discriminant that provides a good separation between b jets and
jets of other flavors.
Secondary-vertex reconstruction is performed with an inclusive vertex search amongst the tracks associated to a jet~\cite{bib:AVF}.

Simulated samples of signal and background events were produced using
various event generators and including pileup events. The CMS detector
response is modeled with {\GEANTfour}~\cite{Agostinelli:2002hh}.
The MSSM Higgs signal samples, $\Pp\Pp\rightarrow\bbbar\phi$+X, $\phi\rightarrow\bbbar$, were produced with
{\PYTHIA} v6.424~\cite{Sjostrand:2006za}, which yields the $\pt$ and $\eta$ distributions of the leading associated b jet in good agreement with the NLO calculations~\cite{Dittmaier:2012vm}. The Quantum Chromodynamics (QCD) multijet background events were produced
with {\PYTHIA} and {\ALPGEN}~\cite{Mangano:2002ea}, while for $\ttbar$ + jets events
the {\MADGRAPH}~\cite{madgraph} event generator was used. The next-to-leading order generators are interfaced with {\PYTHIA}.
For all generators, fragmentation, hadronization, and the underlying event are modelled using {\PYTHIA} with tune Z2. The parton density functions (PDF) from CTEQ6L1~\cite{CTEQ} are used.

\section{All-hadronic signature}\label{sec:FullHad}

We search for the Higgs boson in events where the three leading jets
are all b-tagged. A signal would be identified as a peak in the invariant mass
distribution of the two leading jets. Events in the data with only two b tags
among the three leading jets are used to model the background, after proper reweighting, as described
in Section~\ref{subsec:BackgroundModel}.

\subsection{Trigger and event selection}
The large hadronic interaction rate at the LHC poses a major challenge
for triggering.
Events are accepted if
either two or three jets are produced in the pseudorapidity range $|\eta| < 2.6$ and have $\pt$ above certain thresholds.
Due to the increase in instantaneous luminosity as the run progressed the jet triggers had to be changed. Thus the data is divided into three categories. The first (second)
category is characterized by dijet triggers in which the leading jet is required to have
$\pt>46\,(60)\GeV$, and the next-to-leading
jet $\pt>38\,(53)\GeV$.  The third category is similar
to the first but requires a third jet with $\pt>20\GeV$.
The online identification of b jets is performed
by an algorithm based on the impact parameter
significance of the second most significant track associated to the jet as
the b-tagging discriminant.
Only events with at least two jets
passing the online b-tagging requirement are accepted by the trigger.

The triggers with lower thresholds allow for a better exploration of the low mass region, albeit with smaller integrated luminosity.
The inclusion of the higher-threshold triggers allows higher integrated
luminosity, but with the adjusted analysis requirements only the medium to
high mass region can be covered. For this
reason two analysis scenarios are defined: in the low-mass scenario ($M_\phi < 180$\GeV),
events accepted by the low jet $\pt$ threshold triggers (first and third categories) are selected
corresponding to an integrated luminosity of 2.7\fbinv. In the medium-mass scenario ($180 \leq M_\phi \leq 350$\GeV),
a combination of dijet triggers with low and high jet $\pt$ thresholds (first and second categories)
forms an event sample with an integrated luminosity corresponding to 4.0\fbinv.

Events are required to have at least three reconstructed jets with
$|\eta| < 2.2$, where the b-tag efficiency and mistag probability are essentially constant.
The three leading jets must also pass the $\pt$ cuts of 46, 38 and 20\GeV
(60, 53 and 20\GeV), respectively, in the low- (medium-) mass scenario.
A minimal separation of $\Delta R>1$ between the two leading jets, where $\Delta R=\sqrt{(\Delta\eta)^2+(\Delta\phi)^2}$ and
$\Delta\eta$ and $\Delta\phi$ are the pseudorapidity and
azimuthal angle differences between the two jets, is
required to suppress background from gluon splitting to a b-quark pair.

We define a ``triple-b-tag'' sample to search for a signal by requiring
all three leading jets to pass a tight CSV b-tagging
selection requirement, consistent with the online b-tagging demand, at a working point characterized by a misidentification probability
for light-flavor jets of about 0.1\% at an average jet $\pt$ of 80\GeV. The average b-tagging efficiency for true b jets is about 55\% for jets with $80 < \pt < 120$\GeV.
The total numbers of events passing the trigger and offline selections are
106\,626 and 89\,637 for the low- and medium-mass scenarios,
respectively. The efficiency of the trigger for signal events passing
the offline selection is 47--67\%, for a Higgs boson mass in the range of 90--350\GeV.

We define a ``double-b-tag'' sample, which is instrumental in estimating
the shape of the background, where only two of the three leading jets have
to pass the above-mentioned criteria, while the remaining untagged
jet does not have to fulfill any b-tagging requirements. Since the double-b-tag sample
is dominated by QCD events with two b jets, it represents a control region suitable
to model the shape of the background contribution.

The secondary-vertex mass, namely the invariant mass calculated from all tracks  forming the secondary vertex,
provides an additional separation between b-, c- and light-flavor jets (attributed to u, d, s, or g partons)
beyond the CSV b-tagging selection requirement. A compact b-tagging
variable for the whole event is constructed
by assigning to each selected jet $j$, where $j$ is the rank of the jet
in the order of decreasing \pt,
 an index $B_j$ which can take one of three possible values. For jets with no reconstructed secondary vertex, or where the secondary vertex mass is below 1\GeV, $B_j$ is set to zero. For intermediate values of the vertex mass between 1--2\GeV the index is set to 1, and for vertex mass larger than 2\GeV it is set to 2.
The three indices $B_1$, $B_2$, and
$B_3$ are combined in an event b-tag variable $X_{123}$, which is defined as follows:
$X_{123} = X_{12} + X_3$, where $X_{12} = 0,$ 1 or 2 depending
on whether $B_1+B_2 < 2$, $2 \leq B_1+B_2 < 3$ or $B_1+B_2
\geq 3$, respectively, and $X_{3}=0$ if $ B_3 < 2$, and $X_{3}=3$
otherwise.

By construction, the event b-tag variable $X_{123}$ can have six possible values
ranging from 0 to 5. The intention of this mapping is to have each
bin populated with sufficient statistics. For event types with a strong triple b-tag signature,
the $X_{123}$ distribution typically shows peaks at values of 2 and 5.

\subsection{Background model and signal extraction}
\label{subsec:BackgroundModel}
The dominant background comes from Quantum Chromodynamics (QCD) multijet production with two or
more jets containing b hadrons, and can neither be fully reduced by
kinematic selection, nor reliably predicted by Monte Carlo (MC) simulation. For this
reason, a method based on control data samples, similar to the one used
in Ref.~\cite{Aaltonen:2011nh} is applied. The background model is
constructed from templates that are derived from the double-b-tag
sample.

We divide the events in the double-b-tag sample into the following categories:
bbx, bxb, and xbb, depending on the rank, sorted by \pt, of the
untagged jet, which is represented by the lower-case letter x. The ranking in descending $\pt$
of the three jets is incorporated in the nomenclature adopted
here, \eg bbx means a sample of events where the two leading jets
are b tagged and the third jet is the untagged jet. The
\textit{true} flavor of the untagged jet can be either light (u,d,s flavor quark, or g, denoted collectively by q), charm (c) or bottom (b).

From these three double-b-tag categories, nine
background templates are constructed by weighting each untagged jet
with the b-tagging probability assuming that its true flavor corresponds to
either a light parton (u, d, s, or g, denoted by Q), a charm (C) or a bottom (B) quark. The convention
is that the capital letter indicates the {\em assumed} flavor of the untagged jet.
The
b-tagging probability for each flavor is determined as a function of jet
$\pt$ and $\eta$ with simulated multijet events. Data/MC scale factors
for the b-tagging efficiencies of b, c, and light flavor jets are
applied where appropriate~\cite{CMS-PAS-BTV-12-001}.

Each background template is a distribution in the two-dimensional space
spanned by $M_{12}$, the dijet
mass of the two leading jets, and the event b-tag variable $X_{123}$.

The following nine background templates are thus created:
Qbb; Cbb; Bbb; bQb; bCb; bBb; bbQ; bbC; and bbB.
In the bbb background events, two \bbbar pairs are present.
As pointed out in Ref.~\cite{Aaltonen:2011nh},
the template bbB models mainly bbb events in which the two leading b quarks originate
from the same \bbbar pair in the event, while Bbb and bBb are important to
cover cases where the two leading b quarks originate from different
\bbbar pairs.

The $X_{123}$ dimension of the templates is modeled in a similar way. Each of the three
possible values of the secondary-vertex mass index of the untagged jet
is taken into account with a weight according to the probability that a jet
will end up in a given bin of the secondary-vertex mass distribution. These probabilities,
parametrized as a function of the jet $\pt$ and $\eta$, have been determined for
each flavor using jets from simulated \ttbar events.

\begin{figure}[!ht]
\centering
\includegraphics[width=\cmsFigWidthHalf]{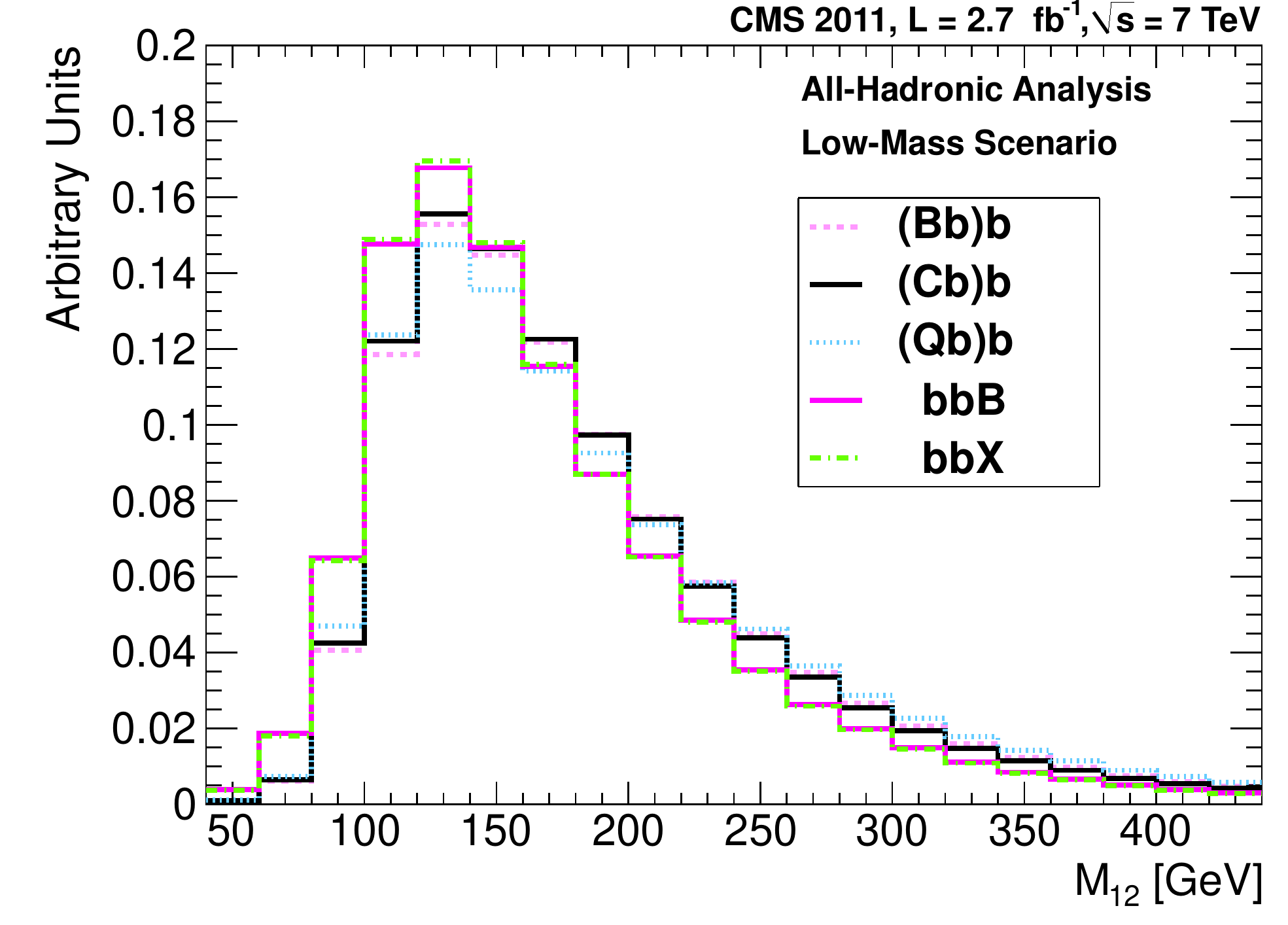}
\includegraphics[width=\cmsFigWidthHalf]{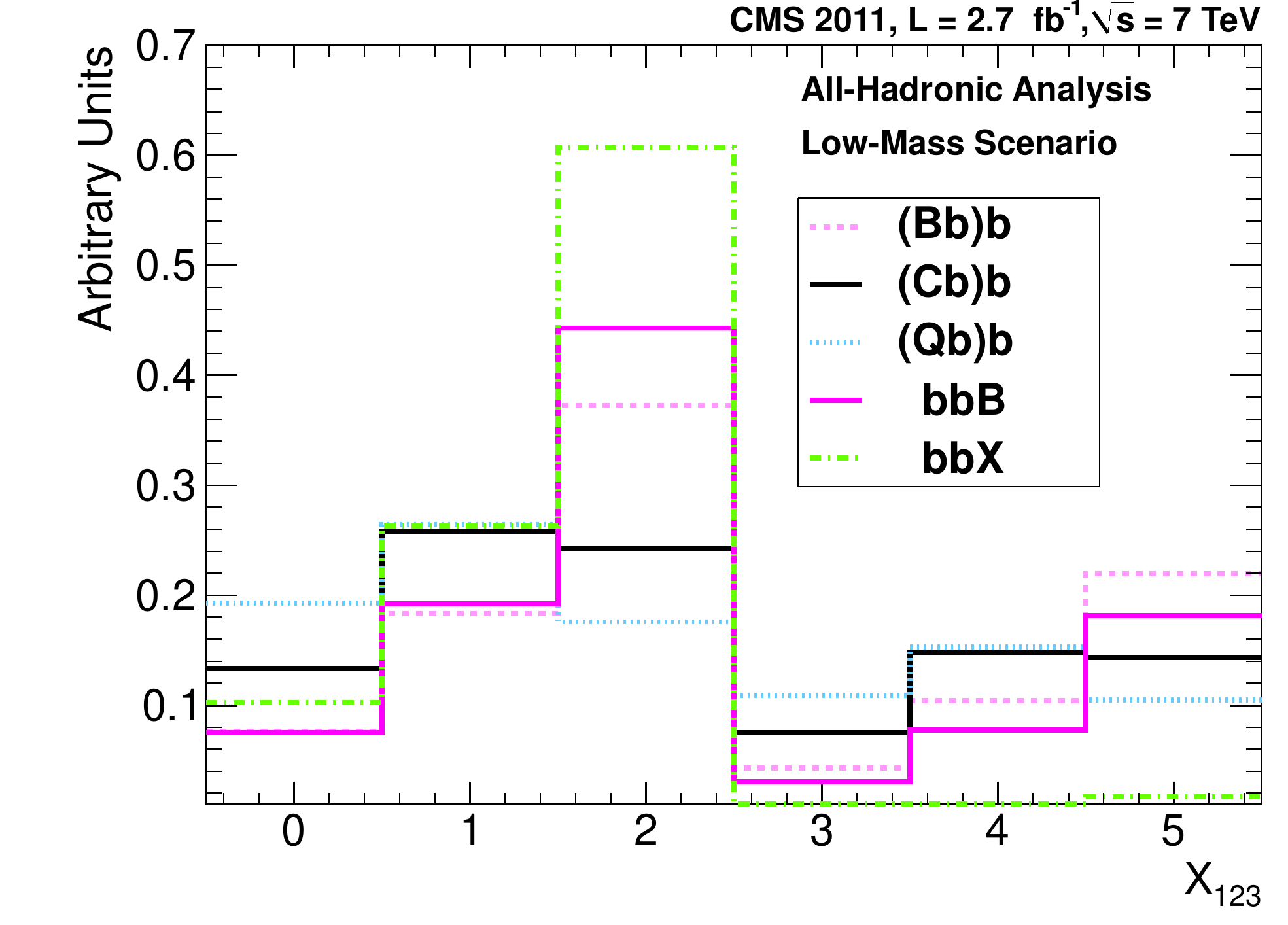}

\caption{The $M_{12}$ (\cmsLeft) and $X_{123}$ (\cmsRight) projections of the five background templates,
(Bb)b, (Cb)b, (Qb)b, bbB, and bbX, for the low-mass scenario.}
\label{fig:templates}
\end{figure}

Some of the nine templates are similar to each other in shape both for $M_{12}$ and $X_{123}$.
In the cases where one of the two
leading jets is untagged, e.g. Qbb, and bQb, the
templates are combined, resulting in a merged template
(Qb)b = Qbb + bQb. By analogy, also (Cb)b and (Bb)b are obtained.
When the third-leading jet is the untagged one and the assumptions of its
flavor are either Q or C, the bbQ and the bbC templates
are combined to form the template bbX. The total number of templates
to be fitted to the data is therefore reduced from nine to five, namely (Bb)b, (Cb)b, (Qb)b, bbB and bbX.
The projections of the $M_{12}$ and $X_{123}$ variables are shown in
Fig.~\ref{fig:templates} for the five background templates and for the low-mass scenario.

Templates whose dijet mass spectra resemble each other can
be clearly distinguished with the introduction of the event b-tag variable $X_{123}$. This
is the case for example between (Bb)b and (Cb)b. In general, the event b-tag significantly improves the discrimination among all flavor components modeled.

The background templates whose projections are shown in
Fig.~\ref{fig:templates}
include two additional corrections. The basic assumption of
the background model, that the double-b-tag sample (bb) consists
entirely of events with at least two genuine b jets, is only
approximately correct. Although the remaining contamination from
non-bb events is indeed very small, the impact of the b-tagging
selection could lead to distortions of the background model
and a correction must be applied. This contamination is estimated directly
from the data using a negative b-tagging discriminator~\cite{CMS-PAS-BTV-11-004} constructed with a track-counting algorithm based on the negative impact parameter of the tracks, ordered from the most negative impact parameter significance upward.
The set of events in the double-b-tag sample in which at least one of the b-tagged jets passes a
certain threshold of the negative b-tagging discriminator is used as a model
for the contamination by non-bb events. The threshold is calibrated as a function
of jet $\pt$ with simulated multijet events, such that the
 negative tag rate equals the mistag rate. With this method,
the non-bb contribution is found to be at the level of
3--4\%. This correction results in only a marginal change in
template shape. A second correction is
necessary because the online b-tagging patterns differ in  the double-
and the triple-b-tag samples. The correction is determined from
simulation, and is applied by
appropriate weighting of the events in the double-b-tag sample.

A signal template is obtained for each considered value of the Higgs boson
mass by performing the full selection on the events of the corresponding simulated signal sample.
The mass resolution for combinations where both b jets stem from the Higgs decay ranges from 12--14\% over the mass range
of 90--350\GeV. In addition, combinations with at least one of the b-tagged jets originating from other sources contribute
to the signal mass spectrum. The fraction of true combinations within 1$\sigma$ of the mass resolution increases from 50--90\%.
Similar figures apply also for the semileptonic signature discussed in Section~\ref{sec:SemiLep}.

 The signal is extracted by fitting a linear combination of signal and
 background templates, $N_{\textrm{bbb}}\ ( f_{\text{sig}} T_{\text{sig}} + \sum_{i} f_{\text{bgd}}^{(i)} T_{\text{bgd}}^{(i)} )$, to
 the observed histogram in $M_{12}$ and $X_{123}$ space, where
 $N_{\textrm{bbb}}$ is the total number of selected triple-b-tag events, $T_{\text{bgd}}^{(i)}$
 and $T_{\text{sig}}$ are the above-mentioned background and signal templates,
 each normalized to unity, and $f_{\text{bgd}}^{(i)}$ and $f_{\text{sig}}$ are the background and
 signal fractions determined by the fit. The results of the fit are discussed in Section~\ref{sec:Results}.

\section{Semileptonic signature}\label{sec:SemiLep}

In the semileptonic signature, as for the all-hadronic one, a signal is
searched for in events with three identified b jets, as a peak in the invariant
mass distribution of the two leading jets.  The expected background
distribution and normalization is built using the same distribution for events
with three jets of which only one or two are tagged as b jets, reweighting the
events with a probability derived from a control region and computed with two
different techniques.
The muon requirement in the final state reduces the
absolute signal efficiency, since it selects events where at least one of the b quarks
decayed semileptonically in the muon channel, but it helps to reduce
the event rate at the trigger level, allowing for a lower threshold for the jets.

\subsection{Trigger and event selection}\label{sec:SLSelection}
The data used in the semileptonic analysis were collected using different trigger selections,
to cope with the increasing luminosity.
All the triggers required a muon with a $\pt>12\GeV$ threshold and the presence
of one or two central jets ($|\eta|< 2.6$ ) with transverse momentum above a
given threshold (20 or 30\GeV, depending on the data-taking period).
Furthermore, one or two b-tagged jets are required online. Initially the
track with the second-most significant
impact parameter was used. Later, when a
second online b-tag was introduced, the selection was on the first track, in
order to retain enough signal efficiency even with this tighter selection.
An integrated luminosity of 4.8\fbinv has been analyzed, and
about $1.67\times{10}^{7}$ events were collected.

The offline analysis requires a muon
with $\pt>15\GeV$, at least three jets
with $|\eta|<2.6$, having transverse momentum $\pt>30\GeV$ for the first two
and $\pt>20\GeV$ for the third one. The separation between any
pair of jets has to be
$\Delta{R}>1$.
The two leading jets must be b-tagged using the CSV b-tagging algorithm with a
working point giving mistag probability for light jets
of about 0.3\%.
The muon must be contained in one of the two leading jets.
The final selection for the signal search adds the requirement that the third
jet is also b-tagged, with a looser CSV b-tagging selection requirement, corresponding to a mistag probability of about 1\%.
The total number of events which pass the selection is 60\,195.

The relative efficiencies of the triggers with respect to the offline
selection criteria were measured using lower-threshold single-muon
triggers. These efficiencies are found to be about 45--60\%,
depending on the Higgs boson mass and the trigger.

\subsection{Background determination and signal extraction}\label{sec:SLBackground}
As in the all-hadronic final state, the major backgrounds for the semileptonic
final state are multijet events from  hard-scattering processes.
Other background processes, such as
$\ttbar+\text{jets}$ and $\cPZ\to\bbbar+\text{jets}$, are
predicted by the MC simulation to be less than 1\% of the total background.
Other possible backgrounds from events with multiple vector
bosons (ZZ, ZW, WW) are negligible.

Two methods, both derived from data, have been developed to predict the expected background.
The first is based on the computation of b-tagging
probabilities of the third jet; and the second is based
on a nearest-neighbor-in-parameter-space technique.
They are able to predict the yield
and shape of the multijet background as well as other minor contributions.
The two methods use completely exclusive data samples, so their two predictions
are independent. The first method uses double-b-tag samples (bbj) and the
second uses single-b-tag samples (bjj) with the double-b-tag events removed.

Both methods require a background-rich sample to serve as a control region.
We construct a discriminating variable with a likelihood ratio, using various
kinematic inputs: the \pt of the b jets; separation in $\phi$ and $\eta$
of the b jets; separation in $\phi$ and $\eta$ between the third jet and the combination of the two
leading jets; and the b jet multiplicity.
Two versions of this discriminating variable are used: one for the  low-mass region
($M_\phi\le180\GeV$) and another for the medium-mass region ($M_\phi>180\GeV$).
For both mass ranges, the control region is defined as the sample of events
having a low value for the discriminating variable, where the background is
enriched and the signal depleted, and the signal region is defined as the
complementary sample.
The number of events in the signal region is 33\,366 and 16\,866, for the low- and medium-mass regions, respectively.

The first method, henceforth called the matrix method, uses the probability that a jet is
identified as a b jet to predict the background in the signal region.
The predicted distribution of any observable $x$
in the three-b-jet sample can be calculated by rescaling, on an event-by-event basis, the
same distribution for the bbj sample by the probability
$P_\textrm{b}$ of the third jet to be b-tagged.
Taking into account the contribution of b, c, and light jets, the probability expands as:
$P_\text{b}=\epsilon_\text{b}\cdot{f_\text{b}}+\epsilon_\text{c}\cdot{f_\text{c}}+\epsilon_\text{q}\cdot{\left(1-f_\text{b}-f_\text{c}\right)}$,
where $\epsilon_\text{b, c, q}$ are the
probabilities (or b-tagging efficiencies) for a jet to be b-tagged, if it is
originated from b, c, or light parton, respectively.  The $f_\text{b, c}$ are the
probabilities that the third jet originates from the corresponding quark, which
depend on the kinematics of the event and of the third b jet.

The b-tagging efficiencies are taken from the MC simulation and checked with several
methods derived from data~\cite{CMS-PAS-BTV-11-001}.
Data/MC scale factors (close to unity within a few percent) are applied for the efficiencies and
the corresponding uncertainties used as systematics uncertainties.
The efficiencies are parametrized as functions of the third jet
\pt, $\eta$ and charged-particle multiplicities.
The quark flavor fractions are obtained directly from data by a simultaneous fit of two flavor-sensitive
observables, using templates built from simulated events with b, c and
light quarks.
The first variable used is a b-tag discriminator which uses the confidence level that the four tracks with the highest impact parameter in the jet are consistent
with originating from the primary vertex.
The second is the invariant mass associated with the secondary vertex, if it has been
reconstructed.
The parametrization for quark fractions also depends on the angular separation between the
three b jets.
Only events in the control region are used to obtain the quark
fraction, which is then used to predict the background in both control and
signal region.

The second method, called the nearest-neighbor method, exploits the fact that the probability for an
event to appear signal-like depends on several event and jet variables.
Events from the background enhanced control region are categorized according to
several such variables, and are used to create a multi-parameter background
prediction.
The method uses the bjj sample, and determines, for each event, the
probability to pass the final selection.
Starting from the bjj sample, excluding the bbj events, we can identify four
disjoint subsets with which we work: (1) bjj (including bjb) in the control
region, (2) bbb in the control region, (3) bjj in the signal region, and (4) bbb in the signal
region. The sum of the above sets corresponds to the initial sample, where the
leading jet is always b-tagged. We call collectively ``training sample'' the sum
of subsets (1) and (2), and ``testing sample'' the sum of (3) and (4).
The probability that an event in the testing sample passes the full selection is estimated by
considering a larger sample of ``similar'' events in the training sample,
and counting how many of these events pass the full selection.

For each event in the testing sample, referred to as test events, we select
a sample of events with similar kinematics in the training sample.
The probability for a test event to pass the final selection is calculated by
selecting a sample of 100 training events inside a hyper-ellipsoid in the
multi-dimensional space of event and jet observables, centered at the test
event.
The training events are chosen as those having the smallest
multi-dimensional distance $D$, where
$D^{2}=\sum_{i=1}^{n_V}w_i^2(x_i{}^\text{test}-x_i{}^\text{training})^2$,
$x_i$ is the variable defining the test and training event, and $n_V$ is the number of variables.
The weights $w_i$ account for the different dispersions of the variables
and their different sensitivity to the b-tagging probability.
A total of $n_V=14$ different variables are used, including \pt, $\eta$ and the charged-particle
multiplicity of the jets, the angular separation between the jets, and the invariant mass and
transverse momentum of the combined jet-jet system.
The weights $w_i$ are computed from the derivative of the probability
for an event to pass the final selection as a function of the variable $x_i$.
We then compute the numbers of bbb and bjj events inside this training sample.
Finally, the probability for test events to have three b-tagged jets is computed as
the ratio of bbb to bjj events in the training sample, using a weighted
average, with $1/D^2$ as weight. The probability obtained this way is then
applied, event-by-event, to the sample of test events to predict the invariant
mass distribution of the sample passing the final selection.
This method gives a prediction of the background shape
in the signal region independent from that obtained with the matrix method.

\begin{figure}[!h]
\includegraphics[width=\cmsFigWidthHalf]{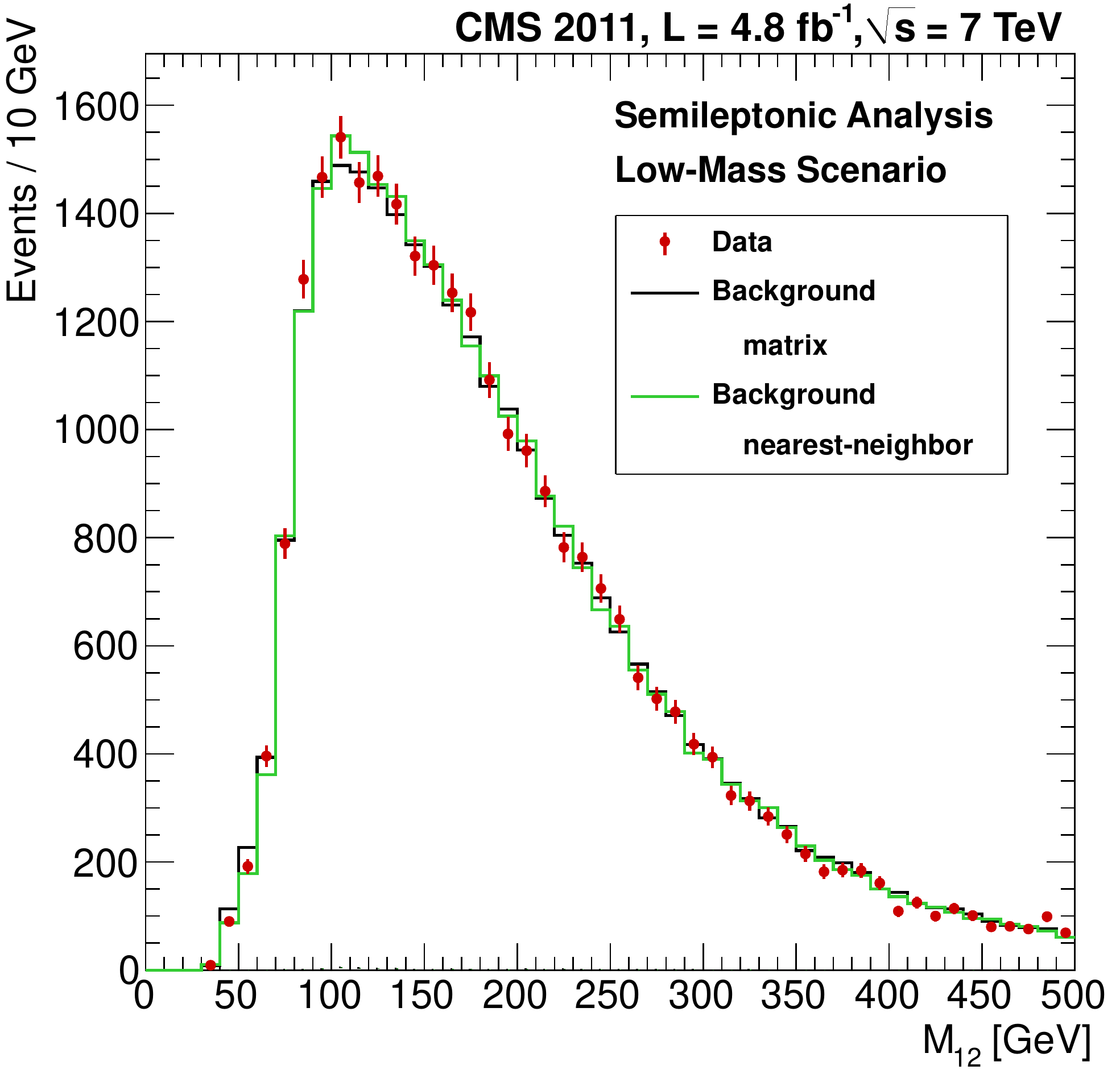}
\includegraphics[width=\cmsFigWidthHalf]{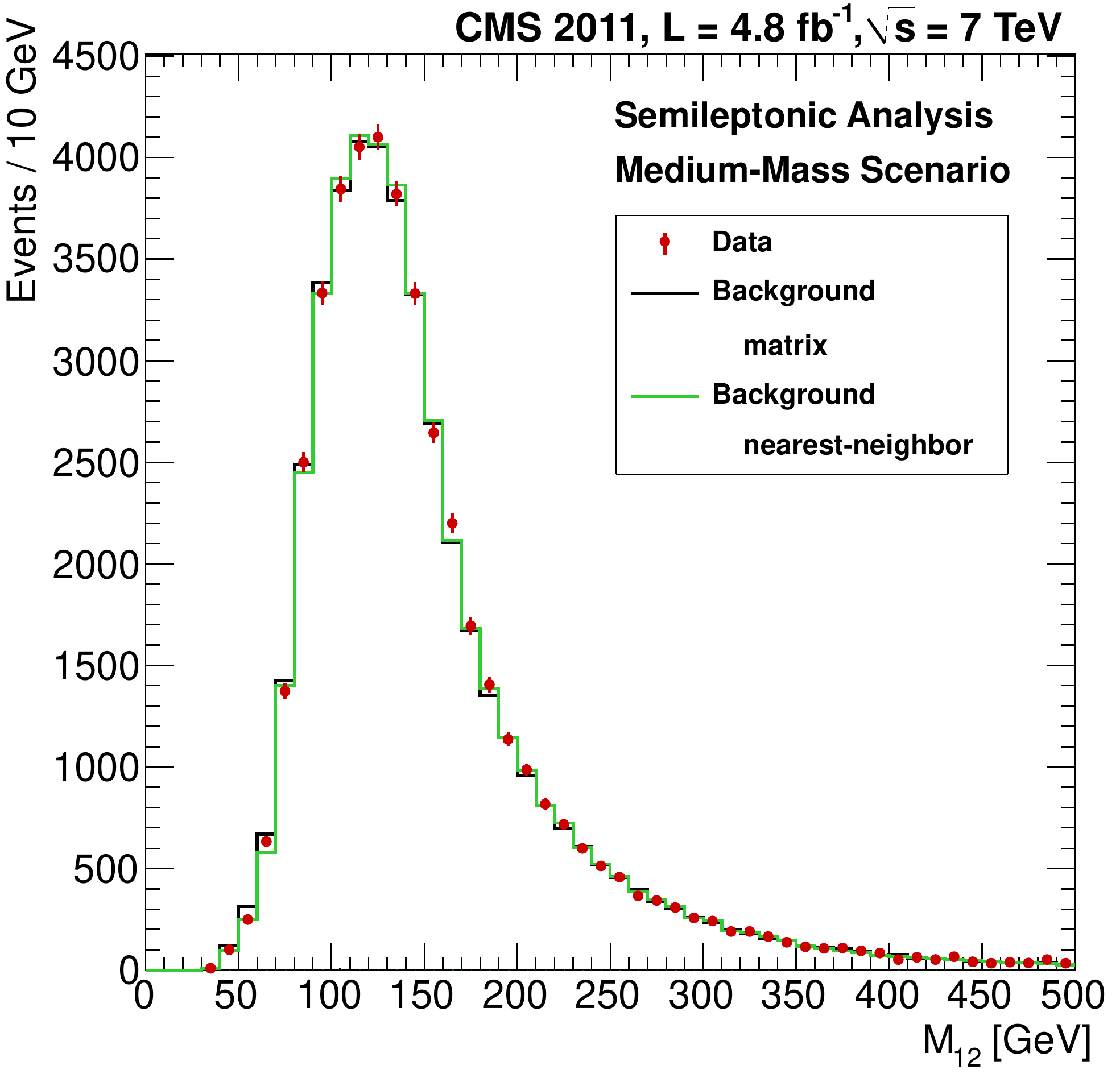}
    \caption{Invariant mass of the two leading jets for data in the
        control regions, for low- (\cmsLeft) and medium-mass (\cmsRight) regions.
        Predictions with matrix (black histogram), with nearest-neighbor methods (green histogram), and data (red
            dots) are overlaid.  The predictions are normalized to the data.}
    \label{fig:bbb:minv01}
\end{figure}

The background predictions for the invariant mass distribution of the two
leading jets from the two methods described above are shown in
Fig.~\ref{fig:bbb:minv01}. They are compared with the actual distribution in
events with three b-tagged jets in the control region (low value of the discriminator), for low- and medium-mass regions.
The predictions are normalized to the number of events seen; the absolute
normalization of the prediction will be discussed in
Section~\ref{sec:Systematics}.

Because the matrix and nearest-neighbor methods use exclusive data samples, we can combine their results.
This is done by performing a weighted average of their bin-by-bin predictions,
using the statistical uncertainties $\sigma_i$ as weights ($w=1/\sigma_i^2$).
In case the $\chi^2$ of the average is greater than 1,
$(\sqrt{\chi^2}-1)\cdot\sigma_i$ is used, bin-by-bin, as an additional
systematic uncertainty, following the Particle Data Group prescription~\cite{ref:pdg}.

\section{Systematic uncertainties}\label{sec:Systematics}

Various systematic uncertainties on the expected signal and
background estimates affect the cross section estimation and, consequently, its
interpretation within the MSSM.  In both analyses the main source of systematic uncertainty
on the estimated signal yield comes from
uncertainties related to jet reconstruction and b tagging. The second source is the
turn-on behavior of the trigger efficiency, given the rather low thresholds used in the event
selection. Other sources include uncertainties on the integrated luminosity and
lepton identification. The theoretical cross sections used for the MSSM interpretation are subject
to factorization and renormalization scale
uncertainties, uncertainties due to the choice of parton distribution functions and $\alpha_s$,
and uncertainties from the underlying event and parton shower modelling~\cite{Botje:2011sn}.
These uncertainties affect only the computation of the upper limits for the MSSM parameter $\tan\beta$ from the cross section results.
The systematic effects directly affecting the signal efficiency, hence the cross section and MSSM interpretation,
are summarized in Table~\ref{tab:Syst}.

\begin{table*}[bt]
    \topcaption{Systematic uncertainties on the signal yield from the various
    sources listed in the first column. The following two columns list the resulting uncertainties
    in the all-hadronic and semileptonic analyses. The upper group is for the
    signal, the lower for the model-dependent limits.
    A range indicates the variation across the probed Higgs boson mass values.
        The source with $^\dagger$ also affects the background, while those with
        $^\star$ only affect the model-dependent results in the space of the MSSM
parameters $\mA$ and $\tan\beta$. The sources labeled with ``rate" affect only
the total signal yield, those with ``shape" also the shape of the signal.
}
    \label{tab:Syst}
    \begin{center}
        \begin{tabular}{l|c c| c}
            \hline
            Source                                       & All-hadronic      & Semileptonic  & Type \\
            \hline
            \hline
            Trigger efficiency                           & $10\%$            & $3-5\%$ & rate\\
            Online b-tagging efficiency                  & $32\%$            & -- & rate \\
            Offline b-tagging efficiency                 & 10--13\%$^\dagger$  & 12\% & shape/rate \\
            b-tagging efficiency dependence on topology  & 6\%             & -- & rate \\
            Jet energy scale                             & 1.4--6.8\%       & 3.1\% & shape/rate\\
            Jet energy resolution                        & 0.6--1.3\%       & 1.9\% & shape/rate \\
            Muon momentum scale and resolution           & --                 & 1\% & rate\\
            Signal Monte Carlo statistics                & \multicolumn{2}{c|}{1.1--2.6\%} &rate\\
            Integrated luminosity   & \multicolumn{2}{c|}{2.2\%} & rate\\
            \hline
            {PDF} and $\alpha_s$ uncertainties     & 3--6\%$^\star$     & 2.7--4.7\%$^\star$ & rate\\
            Factorization and renormalization QCD scale    & \multicolumn{2}{c|}{6--28\%$^\star$} &rate\\
            Underlying event and parton showering          & \multicolumn{2}{c|}{4\%$^\star$} &rate\\
\hline
        \end{tabular}
    \end{center}
\end{table*}

There are systematic uncertainties that affect only the all-hadronic or semileptonic analyses.
In the all-hadronic analysis, Table~\ref{tab:Syst} includes systematic uncertainties
related to the efficiency of the online b-tag selection relative to
that applied offline, and to a slight dependence of the b-tagging
efficiency on the jet topology.
Various uncertainties also affect the shapes of the signal
and background templates used in the fit.
Shape-altering effects from uncertainties on the jet energy scale, jet energy resolution,
b-tagging efficiency and mistag rates are accounted for in the fits with nuisance parameters.
For the background templates, only the latter two are relevant.
In the following we quantify background-related systematic uncertainties by
their effect on the estimated signal fraction $f_{\text{sig}}$ (defined in Section~\ref{sec:FullHad}).
The uncertainty
arising from the jet energy scale and the b-tagging efficiency on the template shape increases the $f_{\text{sig}}$ uncertainty by typically
0.1--0.4\%; the corresponding effect from the jet energy resolution uncertainty is 0.1--0.3\%.
Additional shape-altering systematic uncertainties arise from the impurity
of the double-b-tag sample and the online b-tagging correction to the background templates shape.
The contribution of the former to the $f_{\text{sig}}$ systematic uncertainty ranges between 0.1--0.3\% in the mass
range 90--130\GeV, and is below 0.1\% elsewhere. The effect of the latter correction ranges
from 0.1\% to 0.4\% in the mass range 90--160\GeV, and is below 0.1\% elsewhere. The statistical uncertainty on the offline
b-tagging efficiency values is propagated into the templates and accounted for in the fitting
procedure. The impact on the $f_{\text{sig}}$ uncertainty is typically in the range 0.1--0.6\%.

In the semileptonic analysis there are
uncertainties on both the background shape and normalization.
The shape-related uncertainty is inferred by the comparison of the background predictions obtained
with the two methods described in Section \ref{sec:SLBackground}. The corresponding uncertainty
scaling is included on a bin-by-bin basis in the binned maximum likelihood fit to the
distribution of the final observable. The background normalization uncertainty has two components:
the first is related to the level of agreement between the predicted $M_{12}$ distribution and the
actual bbb one in the data control region and the second is related
to the extrapolation of this prediction from the control region to the signal
region.
The ratio between the predicted $M_{12}$ distribution and the actual
bbb one in the control region as seen in the data is used to
normalize the prediction in the signal region, and its uncertainty is used as a
systematic uncertainty. The scale factor is $0.877\pm0.007$ for the low-mass
region and $0.885\pm0.006$ for the medium-mass region.
For the extrapolation from the control region to the signal region, the MC
simulation shows a constant ratio between the predicted $M_{12}$ distribution and the actual bbb one
in the signal region.  The additional correction is $1.01\pm0.04$ and $1.02\pm0.05$ for the
low- and medium-mass regions, respectively. The uncertainties on these corrections are used as systematic uncertainties for the background
normalization: 4.4\% and 5.0\% for the low- and medium-mass ranges, respectively.

\section{Results}\label{sec:Results}

\begin{figure*}[!ht]
\centering
\includegraphics[width=\cmsFigWidthHalf]{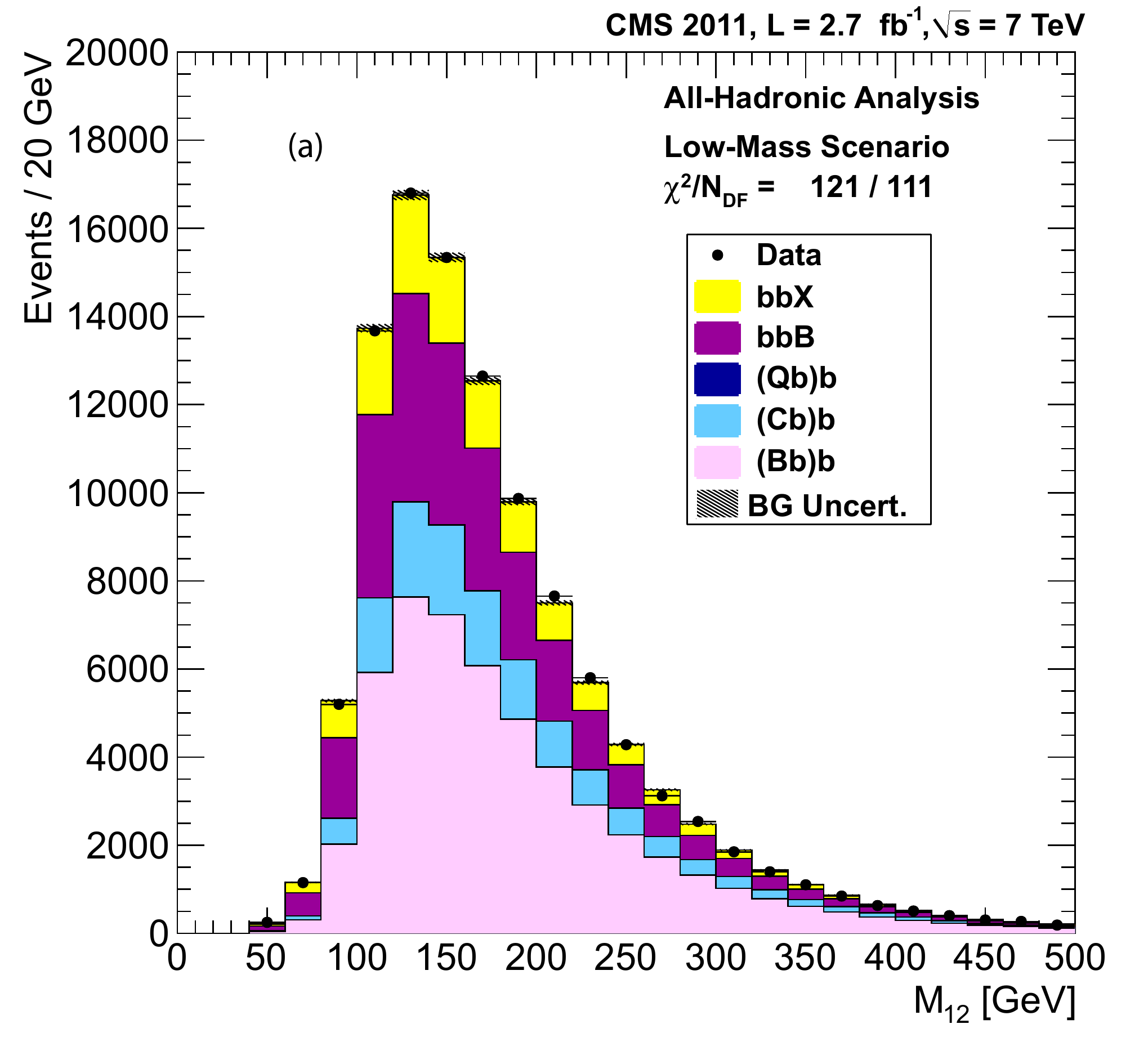}
\includegraphics[width=\cmsFigWidthHalf]{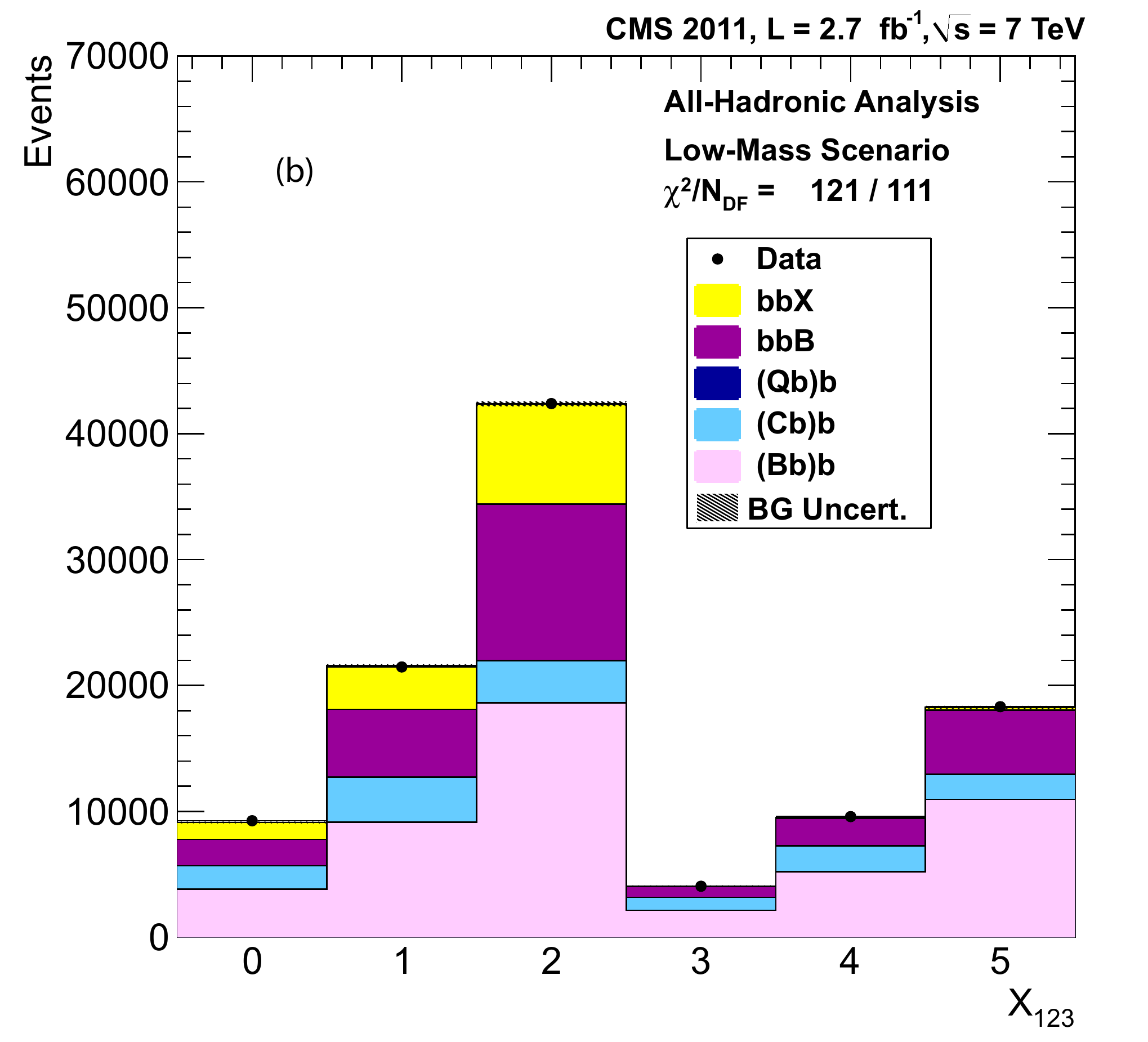}
\includegraphics[width=\cmsFigWidthHalf]{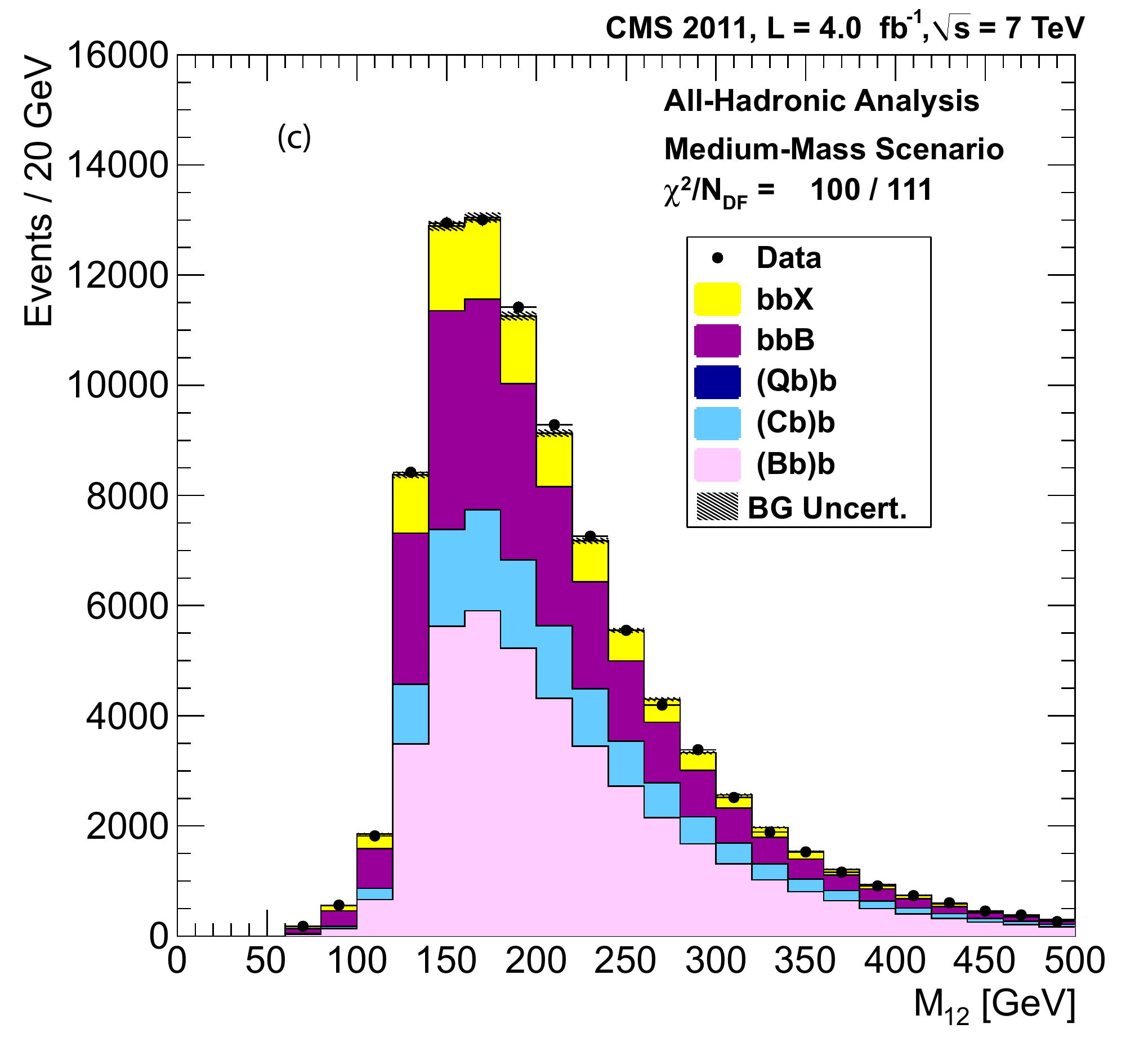}
\includegraphics[width=\cmsFigWidthHalf]{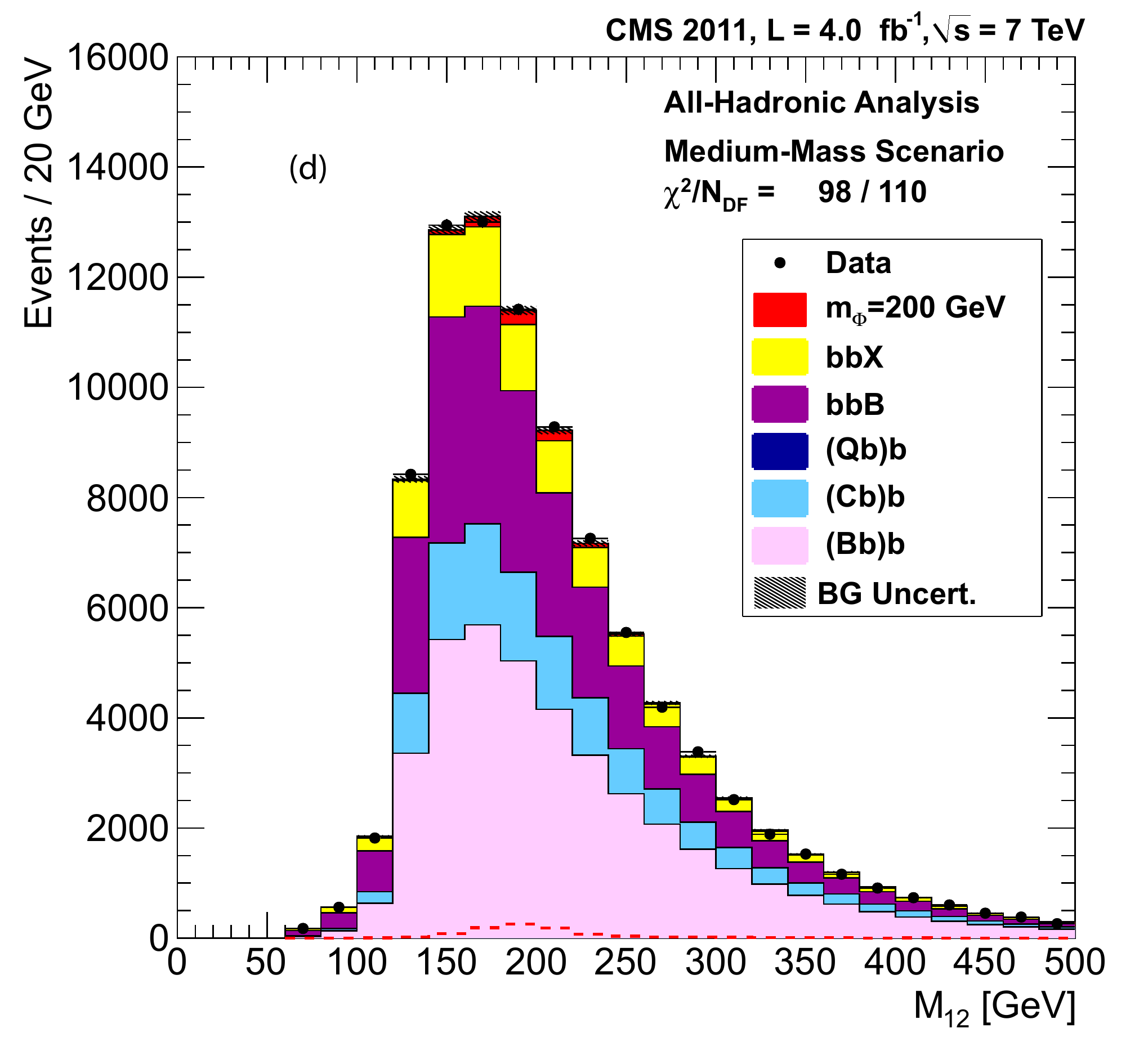}
\caption{Results from the all-hadronic analysis. Top row: Result of the background-only fit in the triple-b-tag
  samples. The plot (a)
  shows the distribution of the dijet mass, $M_{12}$, the plot (b) the distribution
  of the event b-tag variable $X_{123}$ in the low-mass scenario. The hatched area
  at the edge of the summed background histogram corresponds to the uncertainty propagated from the templates. Bottom row: Dijet mass
  distribution in the medium-mass scenario, (d) with the background-only fit,
and (d) including an additional signal template for a MSSM Higgs boson with a mass of
200\GeV. The fitted mass distribution of the Higgs contribution is shown a second time as the dashed histogram at the
bottom of the figure. The fitted contribution of the (Qb)b template is compatible with zero within errors.}
\label{fig:backgroundFit}
\end{figure*}

In the all-hadronic analysis, we first test
the background-only hypothesis by performing a $\chi^2$ fit without
including a signal
template, but only a linear combination of the background templates, as described in Section~\ref{subsec:BackgroundModel}.
The coefficients $f_{\text{bgd}}^{(i)}$ are free parameters, but are constrained to be positive.
Results are shown in
Fig.~\ref{fig:backgroundFit}a-\ref{fig:backgroundFit}c for the low- and medium-mass scenarios. The
background model fits the data well within the uncertainty propagated from the templates
(hatched area). According to the fit, the templates associated with production of three b jets provide the dominant contribution
to the background.

Subsequently, a signal template is included together with the
background templates in the fit, with its fraction $f_{\text{sig}}$ also allowed to vary
freely.  The fit is performed for Higgs boson masses from 90
to 350\GeV. The fit for a Higgs boson mass of 200\GeV
in the medium-mass scenario is illustrated in
Fig.~\ref{fig:backgroundFit}d.

\begin{figure}[htb]
\includegraphics[width=\cmsFigWidthHalf]{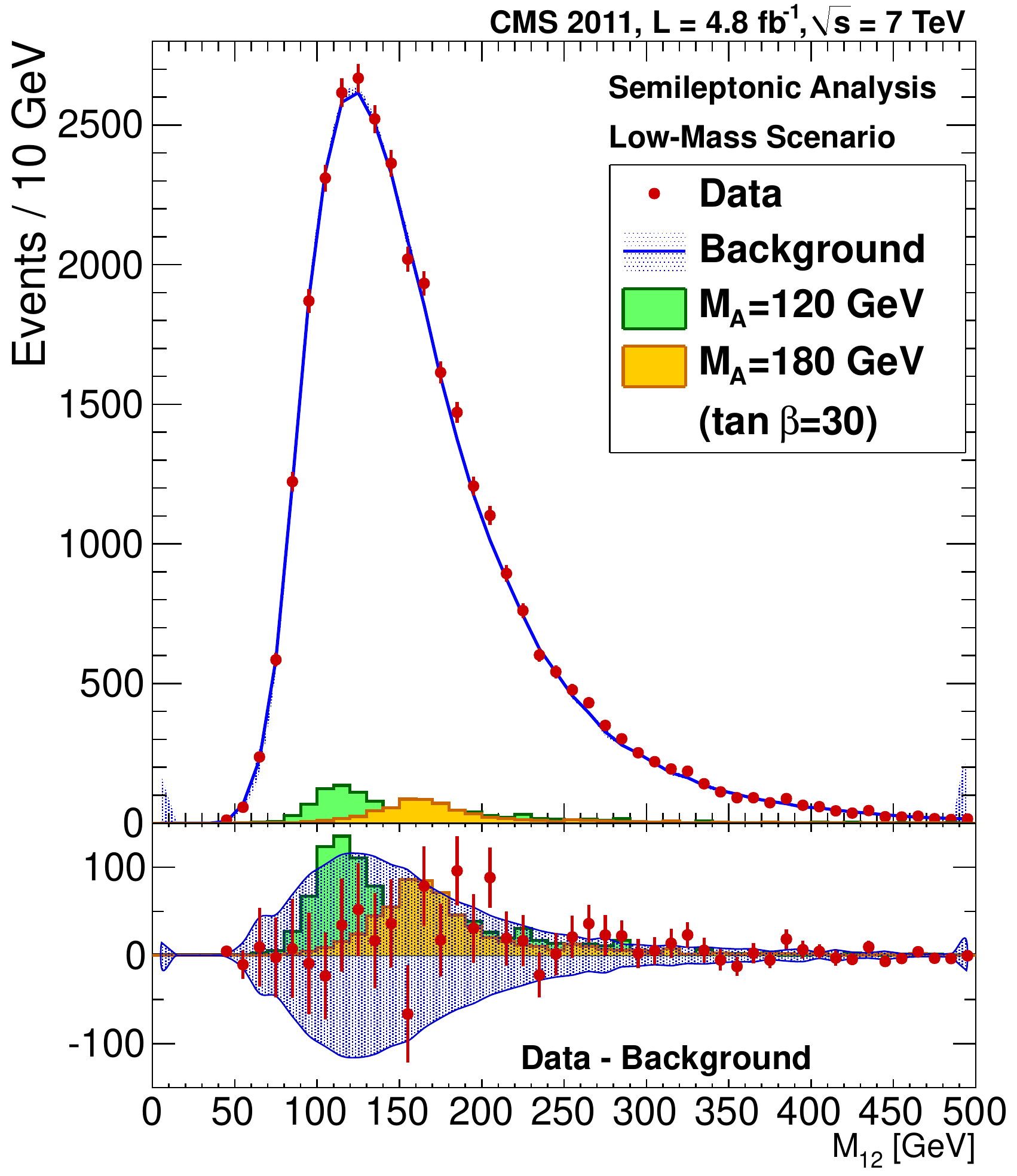}
\includegraphics[width=\cmsFigWidthHalf]{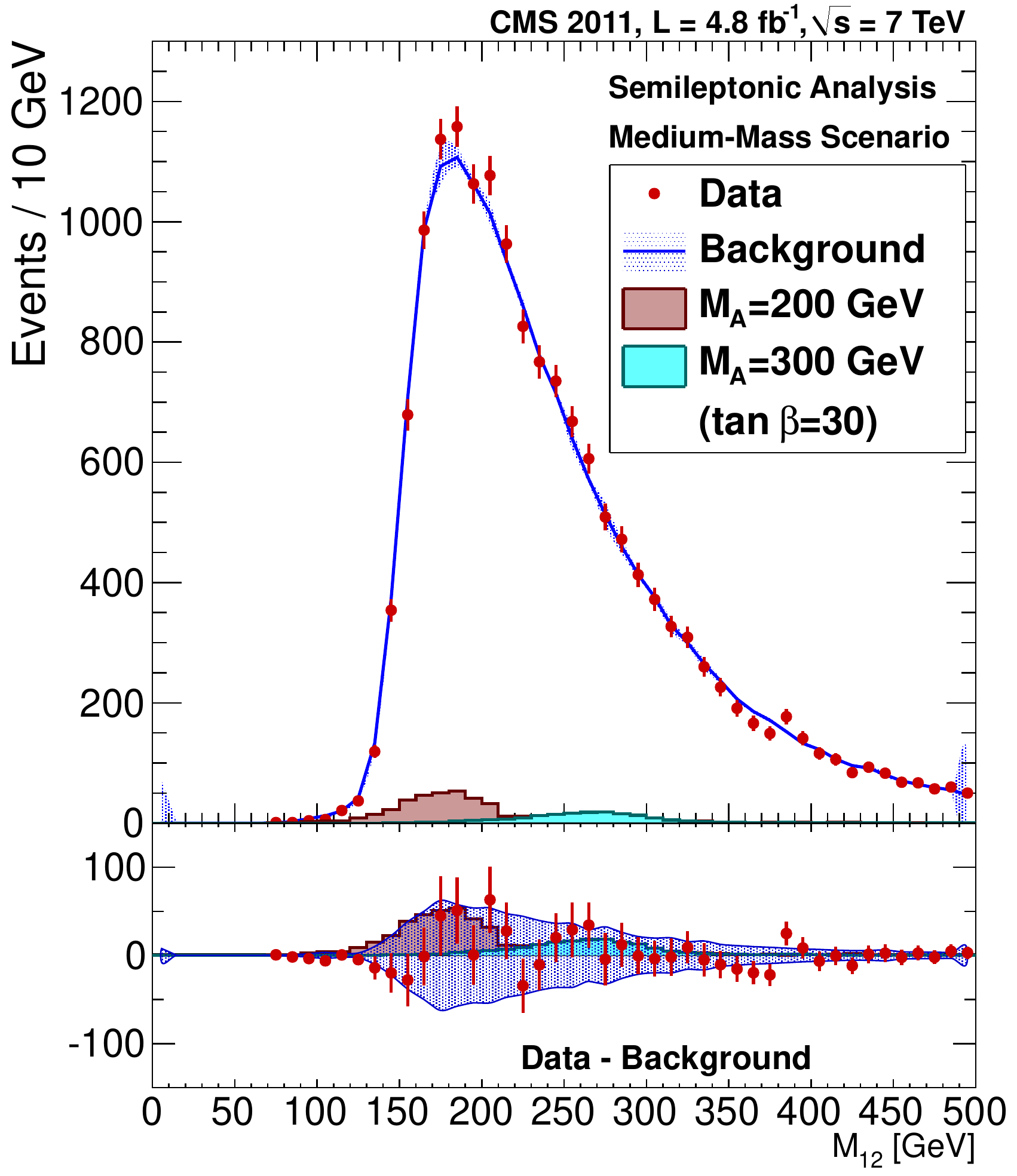}
    \caption{Results from the semileptonic analysis. Data (red) and predicted background (blue) in the signal region,
      for (\cmsLeft) low-mass range (used for $\mPhi\le180\GeV$) and (\cmsRight) medium-mass range
      (used for $\mPhi>180\GeV$); the expected signal for different $\mA$ and for
      $\tan\beta=30$ in the \mhmax scenario, as described in the text, is also plotted. The difference between data and predicted
      background is also shown: the blue area represent the systematic and
      statistical uncertainties on the background prediction.}
  \label{fig:backSignalRegion}
\end{figure}

The semileptonic analysis uses a binned likelihood fit to the
invariant mass distribution of the two leading jets in the event to extract a
possible MSSM Higgs contribution.
Two different background predictions are considered, for the low- and medium-mass
regions, which are fitted separately.
In the fit the shape and normalisation of the background component are constrained through nuisance parameters as explained in section~\ref{sec:Systematics}.
The predicted background is shown in
Fig.~\ref{fig:backSignalRegion}  for
the two mass ranges, together with an expected signal for
two Higgs boson masses at $\tan\beta=30$.

No significant deviation from background is observed in either analysis, and the \CLs~\cite{Cowan:2010js,CLS1,CLS2,CMS-NOTE-2011-005}
criterion is used to combine both results and determine the 95\%
confidence level (CL) limit on the signal contribution
in the data, using the \textsc{RooStats}~\cite{RooStats} package. To avoid correlations,
in the all-hadronic analysis the events common to the
semileptonic case are removed from the
triple-b-tag samples. The fractions of events removed in the all-hadronic data samples are
2.3\% and 2.7\% for the low- and medium-mass scenarios, respectively. The requirement of a muon in the semileptonic analysis and the harder kinematic selections of the all-hadronic analysis are responsible for such small overlap.
Overlapping events in the simulated signal samples are also removed, although they are found to have negligible effect on the shape of the signal templates.

\begin{figure}[htb]
\centering
\includegraphics[width=\cmsFigWidthHalf]{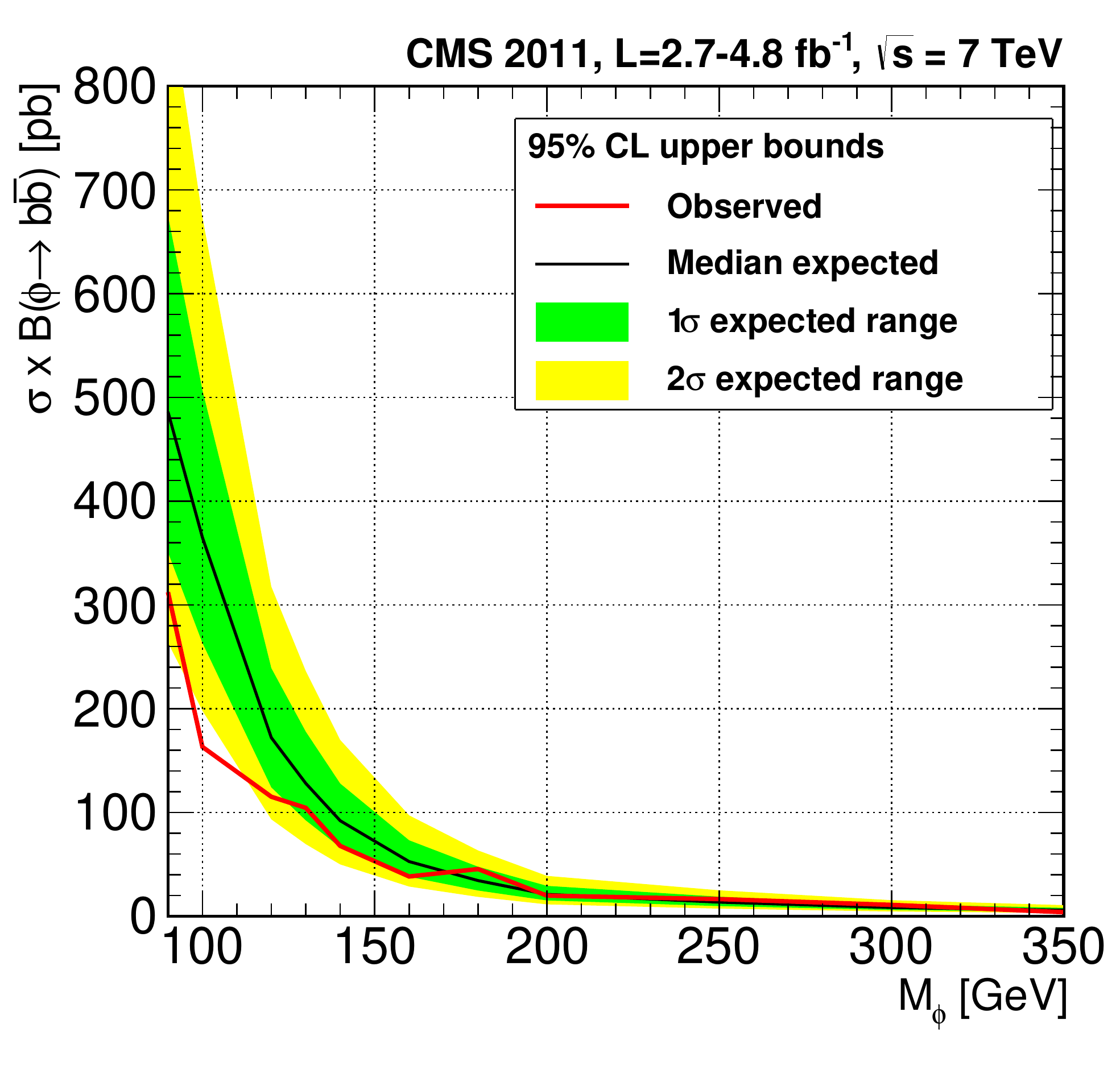}
\includegraphics[width=\cmsFigWidthHalf]{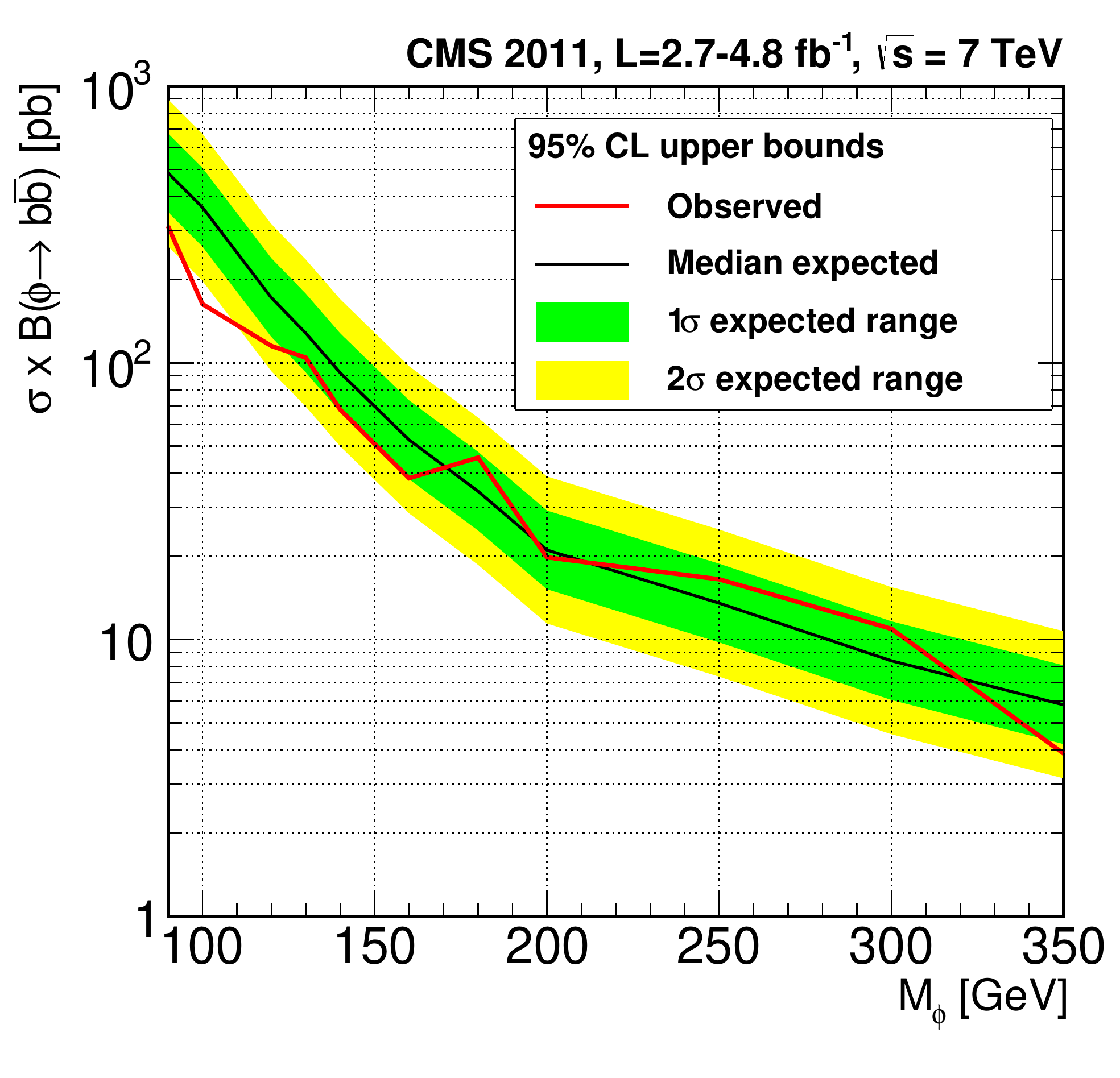}
\caption{Observed and expected upper limits for the cross section times branching fraction at 95\% CL, with linear (\cmsLeft) and logarithmic (\cmsRight) scales, including statistical and systematic uncertainties for the combined all-hadronic and semileptonic results. One- and two-standard deviation ranges for the expected upper limit are also shown.}
  \label{fig:Limit}
\end{figure}

Results are shown graphically in
Fig.~\ref{fig:Limit} in terms of cross section times branching fraction, and reported in Table~\ref{tab:ObsExpXsecLimits}.
There is generally good agreement between the observed and expected upper limits within statistical errors, and no
indication of a signal is seen. The observed upper limits range from about 312\unit{pb} at $\mPhi=90\GeV$ to about 4\unit{pb}
at $\mPhi=350\GeV$. The individual observed limits
of the two signatures are also displayed. The all-hadronic signature
has a generally larger signal efficiency, but requires higher thresholds for jet energies, while
the presence of a muon in the semileptonic signature allows for lower thresholds at the cost of lower
signal efficiency. As a result, both signatures are comparable in sensitivity.

\begin{figure}[htb]
\centering
\includegraphics[width=\cmsFigWidthHalf]{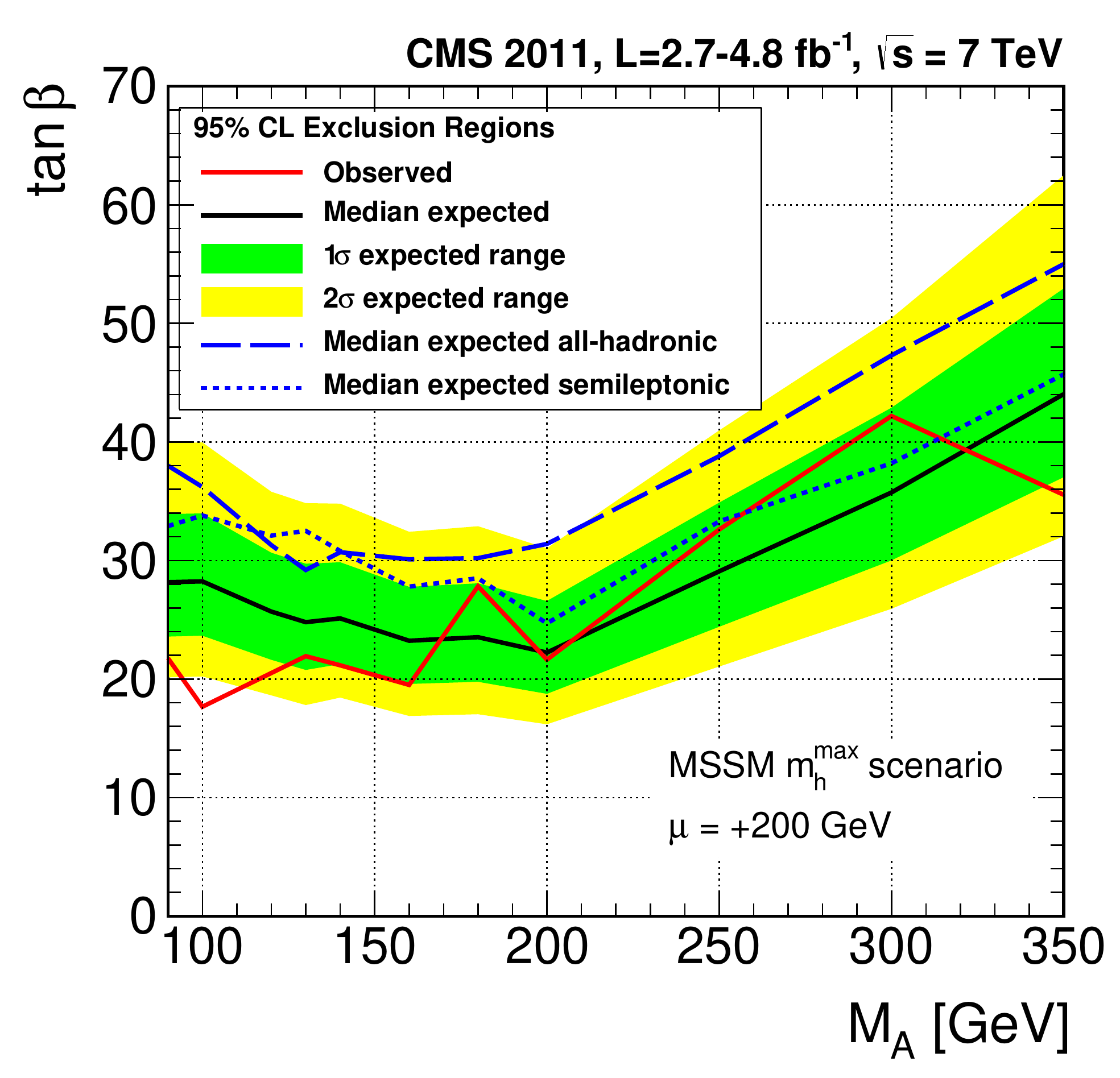}
\includegraphics[width=\cmsFigWidthHalf]{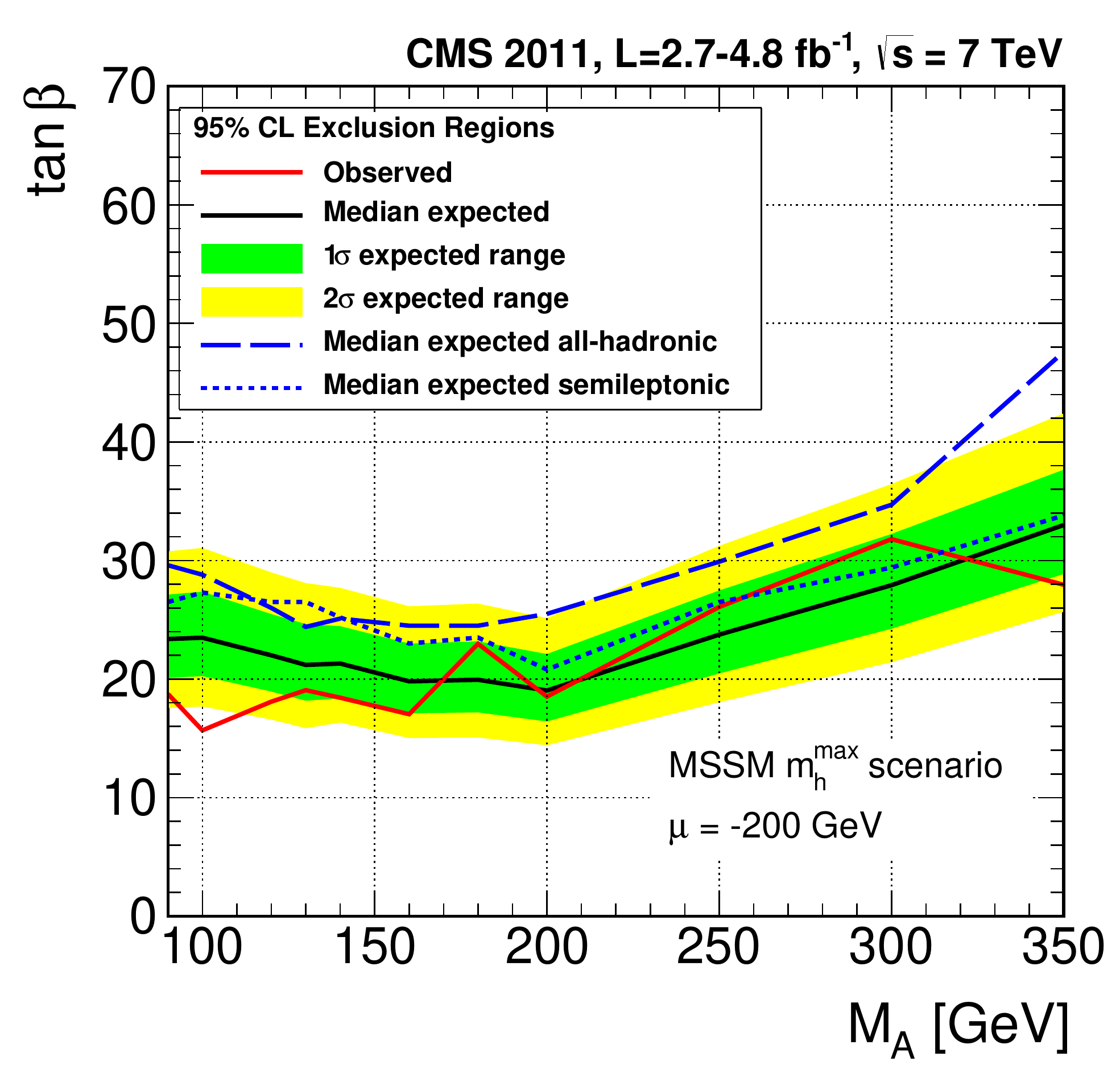}
\caption{Observed upper limits at 95\% CL on $\tan\beta$ as a function of \mA, including the statistical and systematic uncertainties, in the \mhmax benchmark scenario, both for $\mu = +200\GeV$ (\cmsLeft) and $\mu = -200\GeV$ (\cmsRight), for the combined all-hadronic and semileptonic results. One- and two-standard deviation ranges for the expected upper limit are represented by the color bands. The expected upper limits for each of two signatures are also shown (dashed and dotted lines).}
\label{fig:limitTanBMASyst}
\end{figure}

Figure~\ref{fig:limitTanBMASyst} presents the results in the MSSM framework
as a function of the MSSM parameters $\mA$ and $\tan\beta$,
combining the
individual results of the two analyses, including all the statistical and
systematic uncertainties as well as correlations.
We use the MSSM \mhmax benchmark scenario~\cite{ref:mHmax1,ref:mHmax2}, which is designed to maximize the
theoretical upper bound on $\mh$ for a given $\tan\beta$ and fixed $M_\text{SUSY}$.
Even though its parameters are under tension with the latest experimental results~\cite{Carena:2013qia},
it is currently still the most suitable benchmark scenario to compare the sensitivity of different analyses channels.
The definition of theory
parameters in the \mhmax benchmark scenario is the following:
$M_{SUSY}=1\TeV$; $X_\textrm{t}=2M_\text{SUSY}$; $\mu=200\GeV$;
$M_{\PSg}=800\GeV$; $M_2=200\GeV$; and $A_b=A_\cPqt$; $M_3=800\GeV$.
Here, $M_\text{SUSY}$ denotes the common soft-SUSY-breaking squark mass of the third generation;
$X_t=A_\cPqt-\mu/\tan\beta^2$ is the stop mixing parameter;
$A_\cPqt$ and $A_\cPqb$ are the stop and sbottom trilinear couplings, respectively;
$\mu$ is the Higgsino mass parameter; $M_{\PSg}$ is the gluino mass; and $M_2$ is the SU(2)-gaugino mass parameter.
The value of $M_1$ is fixed via the unification relation $M_1=(5/3)M_2\sin\theta_\PW/\cos\theta_\PW$.
The expected cross section and branching fraction, in the MSSM framework,
are calculated by {\sc bbh@nnlo} ~\cite{Harlander:2003ai},
in the 5-flavor scheme, and
\textsc{FeynHiggs}~\cite{Heinemeyer:1998yj,Heinemeyer:1998np,Degrassi:2002fi,Frank:2006yh},
respectively. Exclusion plots for two values of $\mu=\pm200\GeV$ are shown.

\begin{figure}[htb]
\centering
\includegraphics[width=\cmsFigWidth]{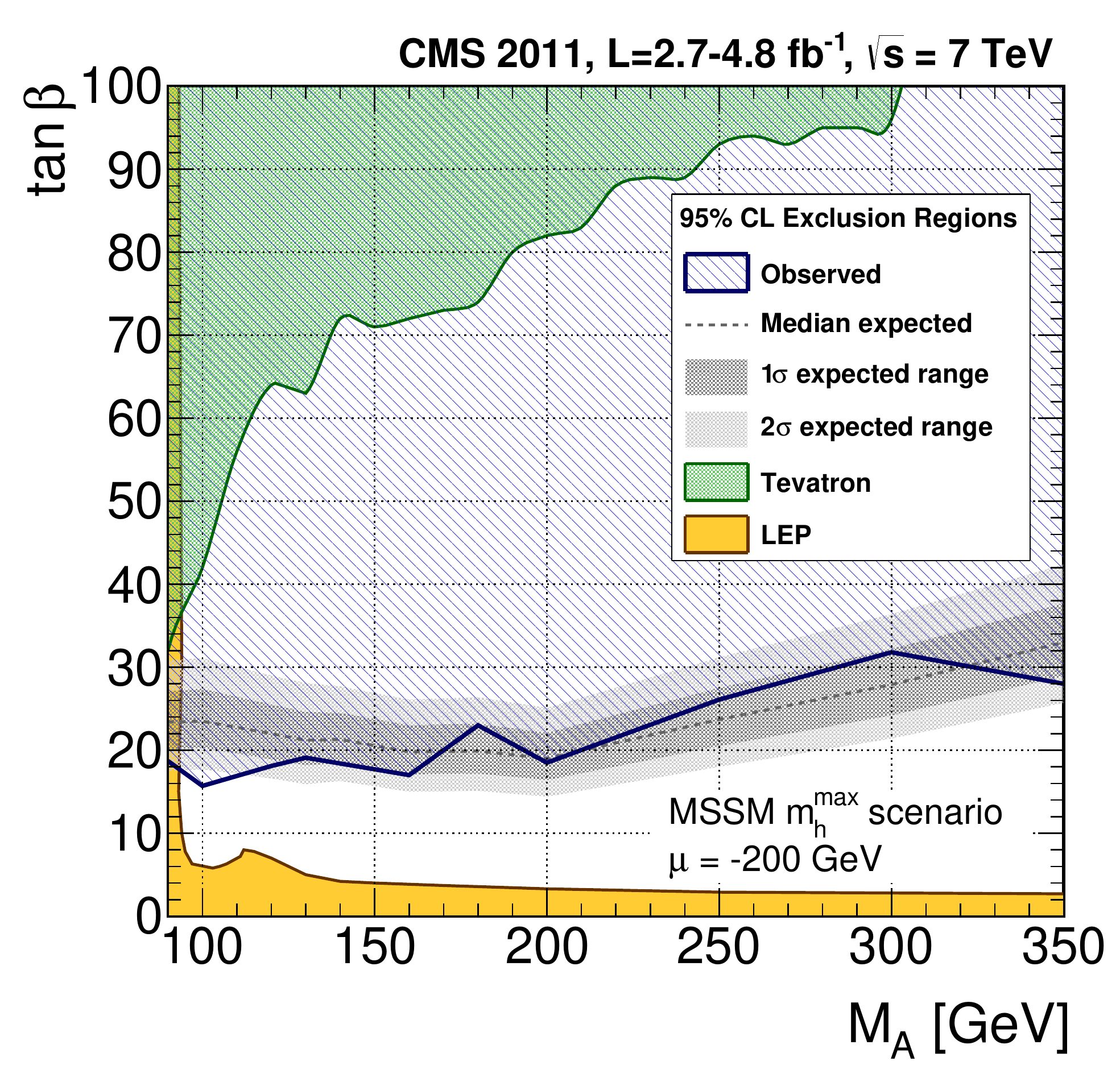}
    \caption{Observed upper limits at 95\% CL on $\tan\beta$ as a function of \mA,
        including the statistical and systematic
      uncertainties, in the \mhmax benchmark scenario with $\mu = -200\GeV$
      for the combined all-hadronic and semileptonic results.
      One- and two-standard deviation ranges for the expected upper limit are represented by the gray bands.
      Previous exclusion regions from LEP~\cite{Schael:2006cr} and Tevatron in the
      multi-b jet channel~\cite{PhysRevD.86.091101} are overlaid.}
  \label{fig:limitTanBMALepTevatron}
\end{figure}

Figure~\ref{fig:limitTanBMALepTevatron} shows the results in
the scenario with $\mu=-200\GeV$, together with previous limits set by
Tevatron~\cite{PhysRevD.86.091101} in the multi-b jet final state, and by LEP~\cite{Schael:2006cr}.
In particular, no excess over the expected SM background is found for high
values of $\tan\beta$ and for a resonance in the 100--150\GeV mass range, as previously
reported by CDF and D0.
The result of this work extends the sensitivity for MSSM
searches in the $\phi\to\bbbar$ decay mode to much lower values of $\tan\beta$, excluding the region where the
excess was reported.

The combined results reported in this Letter, using only the data collected at
the LHC with a center-of-mass of $\sqrt{s}=7\TeV$, provides the most stringent
limits on neutral Higgs boson decay in the \bbbar mode, produced in association with b quarks.

\begin{table*}[tb]
    \topcaption{Expected and observed upper limits at 95\% CL on $\sigma(\Pp\Pp
            \to\cPqb\phi + \textrm{X}) \times {\mathcal{B}}(\phi\to\bbbar)$, in pb, and on $\tan\beta$
        in the \mhmax benchmark scenario for two values of the parameter \mbox{$\mu=\pm200\GeV$}.}
    \label{tab:ObsExpXsecLimits}
    \centering
    \begin{tabular}{c|cc|cc|cc}
        \hline
        & \multicolumn{2}{c|}{$\sigma(\Pp\Pp\to\cPqb\phi + \textrm{X}) \times {\mathcal{B}}(\phi\to\bbbar)$} & \multicolumn{2}{c|}{$\tan\beta$ } & \multicolumn{2}{c}{$\tan\beta$ } \\
        & \multicolumn{2}{c|}{[pb]} & \multicolumn{2}{c|}{($\mu = +200\GeV$)} & \multicolumn{2}{c}{($\mu = -200\GeV$)} \\
        $\mA$(\GeVns{}) & expected & observed & expected  & observed & expected  & observed  \\ \hline \hline
        90            & 486.3  &  312.4   &  28.2     &  21.8    &   23.4   &  18.7  \\
        100           & 365.1  &  163.2   &  28.2     &  17.7    &   23.5   &  15.7  \\
        120           & 172.1  &  115.2   &  25.7     &  20.5    &   22.0   &  18.1  \\
        130           & 128.1  &  104.5   &  24.8     &  21.9    &   21.2   &  19.1  \\
        140           &  92.0  &   67.8   &  25.1     &  21.2    &   21.3   &  18.4  \\
        160           &  52.7  &   38.3   &  23.2     &  19.5    &   19.8   &  17.0  \\
        180           &  34.4  &   45.5   &  23.5     &  27.8    &   19.9   &  23.0  \\
        200           &  21.1  &   19.8   &  22.2     &  21.6    &   19.0   &  18.5  \\
        250           &  13.5  &   16.5   &  29.1     &  32.6    &   23.7   &  26.1  \\
        300           &   8.4  &   10.9   &  35.7     &  42.2    &   27.9   &  31.8  \\
        350           &   5.8  &    3.9   &  44.0     &  35.5    &   33.0   &  28.0  \\ \hline
    \end{tabular}
\end{table*}

\section{Summary and conclusions}\label{sec:Conclusion}

We searched for a Higgs boson decaying into a pair of b quarks,
produced in association with one or more additional b-quark jets.
We used data samples corresponding to an integrated luminosity of
2.7--4.8\fbinv collected in 2011 in proton-proton collisions at a center-of-mass
energy of 7\TeV at the LHC. The data were collected with dedicated multijet triggers
including b-tag selection, utilizing both all-hadronic and
semileptonic event signatures.

The search was performed on a triple-b-tag sample, using the invariant mass of
two leading jets  as a discriminating variable, with a prediction
of the multijet background using control data samples.
The all-hadronic analysis makes
use of a second discriminating variable,
$X_{123}$, that reflects the heavy flavor content of the
event.

No signal is observed above the SM background expectations, and 95\% confidence level upper
limits on the $\Pp\Pp\to\cPqb\phi+\mathrm{X}$, $\phi\to\bbbar$ cross section times branching fraction are derived
in the 90--350\GeV mass range. These results are interpreted,
in the MSSM model and the \mhmax scenario, in terms of bounds
in the space of the parameters, $\mA$ and $\tan\beta$. The 95\% confidence level bound on
  $\tan\beta$ varies from about 18 to 42 in this Higgs boson mass range, thus excluding a region of
parameter space previously unexplored for this final state.

\section*{Acknowledgements}

We congratulate our colleagues in the CERN accelerator departments for the excellent performance of the LHC and thank the technical and administrative staffs at CERN and at other CMS institutes for their contributions to the success of the CMS effort. In addition, we gratefully acknowledge the computing centers and personnel of the Worldwide LHC Computing Grid for delivering so effectively the computing infrastructure essential to our analyses. Finally, we acknowledge the enduring support for the construction and operation of the LHC and the CMS detector provided by the following funding agencies: BMWF and FWF (Austria); FNRS and FWO (Belgium); CNPq, CAPES, FAPERJ, and FAPESP (Brazil); MEYS (Bulgaria); CERN; CAS, MoST, and NSFC (China); COLCIENCIAS (Colombia); MSES (Croatia); RPF (Cyprus); MoER, SF0690030s09 and ERDF (Estonia); Academy of Finland, MEC, and HIP (Finland); CEA and CNRS/IN2P3 (France); BMBF, DFG, and HGF (Germany); GSRT (Greece); OTKA and NKTH (Hungary); DAE and DST (India); IPM (Iran); SFI (Ireland); INFN (Italy); NRF and WCU (Republic of Korea); LAS (Lithuania); CINVESTAV, CONACYT, SEP, and UASLP-FAI (Mexico); MSI (New Zealand); PAEC (Pakistan); MSHE and NSC (Poland); FCT (Portugal); JINR (Armenia, Belarus, Georgia, Ukraine, Uzbekistan); MON, RosAtom, RAS and RFBR (Russia); MSTD (Serbia); SEIDI and CPAN (Spain); Swiss Funding Agencies (Switzerland); NSC (Taipei); ThEPCenter, IPST and NSTDA (Thailand); TUBITAK and TAEK (Turkey); NASU (Ukraine); STFC (United Kingdom); DOE and NSF (USA).
Individuals have received support from the Marie-Curie programme and the European Research Council (European Union); the Leventis Foundation; the A. P. Sloan Foundation; the Alexander von Humboldt Foundation; the Belgian Federal Science Policy Office; the Fonds pour la Formation \`a la Recherche dans l'Industrie et dans l'Agriculture (FRIA-Belgium); the Agentschap voor Innovatie door Wetenschap en Technologie (IWT-Belgium); the Ministry of Education, Youth and Sports (MEYS) of Czech Republic; the Council of Science and Industrial Research, India; the Compagnia di San Paolo (Torino); and the HOMING PLUS programme of Foundation for Polish Science, cofinanced from European Union, Regional Development Fund.

\bibliography{auto_generated}   % will be created by the tdr script.
\cleardoublepage \appendix\section{The CMS Collaboration \label{app:collab}}\begin{sloppypar}\hyphenpenalty=5000\widowpenalty=500\clubpenalty=5000\input{HIG-12-033-authorlist.tex}\end{sloppypar}
\end{document}

%% file: HIG-12-033-authorlist.tex
\textbf{Yerevan Physics Institute,  Yerevan,  Armenia}\\*[0pt]
S.~Chatrchyan, V.~Khachatryan, A.M.~Sirunyan, A.~Tumasyan
\vskip\cmsinstskip
\textbf{Institut f\"{u}r Hochenergiephysik der OeAW,  Wien,  Austria}\\*[0pt]
W.~Adam, E.~Aguilo, T.~Bergauer, M.~Dragicevic, J.~Er\"{o}, C.~Fabjan\cmsAuthorMark{1}, M.~Friedl, R.~Fr\"{u}hwirth\cmsAuthorMark{1}, V.M.~Ghete, N.~H\"{o}rmann, J.~Hrubec, M.~Jeitler\cmsAuthorMark{1}, W.~Kiesenhofer, V.~Kn\"{u}nz, M.~Krammer\cmsAuthorMark{1}, I.~Kr\"{a}tschmer, D.~Liko, I.~Mikulec, M.~Pernicka$^{\textrm{\dag}}$, D.~Rabady\cmsAuthorMark{2}, B.~Rahbaran, C.~Rohringer, H.~Rohringer, R.~Sch\"{o}fbeck, J.~Strauss, A.~Taurok, W.~Waltenberger, C.-E.~Wulz\cmsAuthorMark{1}
% ...
\vskip\cmsinstskip
\dag:~Deceased\\
1:~~Also at Vienna University of Technology, Vienna, Austria\\
2:~~Also at CERN, European Organization for Nuclear Research, Geneva, Switzerland\\
% ...


%% file: HIG-12-033_temp.bbl
\providecommand{\href}[2]{#2}\begingroup\raggedright\begin{thebibliography}{10}%
\makeatletter
\providecommand{\hrefCMSnoop }[0]{\@secondoftwo}%
\makeatother
\providecommand{\doi}{\texttt{doi:}\begingroup \urlstyle{tt}\Url}

\bibitem{Aad:2012gk}
\hrefCMSnoop {} {{ ATLAS} Collaboration, ``{Observation of a new particle in
  the search for the Standard Model Higgs boson with the ATLAS detector at the
  LHC}'',} \textit{ Phys. Lett. B} \textbf{ 716} (2012) 1,
  \href{http://dx.doi.org/10.1016/j.physletb.2012.08.020}{\doi{10.1016/j.physletb.2012.08.020}},
\href{http://www.arXiv.org/abs/1207.7214}{\texttt{ arXiv:1207.7214}}.
%%CITATION = ARXIV:1207.7214;%%.

\bibitem{Chatrchyan:2012gu}
\hrefCMSnoop {} {{ CMS} Collaboration, ``{Observation of a new boson at a mass
  of 125 GeV with the CMS experiment at the LHC}'',} \textit{ Phys. Lett. B}
  \textbf{ 716} (2012) 30,
  \href{http://dx.doi.org/10.1016/j.physletb.2012.08.021}{\doi{10.1016/j.physletb.2012.08.021}},
\href{http://www.arXiv.org/abs/1207.7235}{\texttt{ arXiv:1207.7235}}.
%%CITATION = ARXIV:1207.7235;%%.

\bibitem{Witten-hierarchy}
\hrefCMSnoop {} {E.~Witten, ``Mass Hierarchies in Supersymmetric Theories'',}
  \textit{ Phys. Lett. B} \textbf{ 105} (1981) 267,
  \href{http://dx.doi.org/10.1016/0370-2693(81)90885-6}{\doi{10.1016/0370-2693(81)90885-6}}.

\bibitem{Martin97}
\hrefCMSnoop {} {S.~P. Martin, ``A Supersymmetry Primer'',} (1997).
  \href{http://www.arXiv.org/abs/hep-ph/9709356}{\texttt{
  arXiv:hep-ph/9709356}}. And references therein.

\bibitem{ref:SUSY}
\hrefCMSnoop {} {H.~P. Nilles, ``Supersymmetry, supergravity and particle
  physics'',} \textit{ Physics Reports} \textbf{ 110} (1984) 1,
  \href{http://dx.doi.org/10.1016/0370-1573(84)90008-5}{\doi{10.1016/0370-1573(84)90008-5}}.

\bibitem{ref:Chatrchyan201268}
\hrefCMSnoop {} {{ CMS} Collaboration, ``Search for neutral {H}iggs bosons
  decaying to tau pairs in pp collisions at $\sqrt{s}=7$~{TeV}'',} \textit{
  Phys. Lett. B} \textbf{ 713} (2012) 68,
  \href{http://dx.doi.org/10.1016/j.physletb.2012.05.028}{\doi{10.1016/j.physletb.2012.05.028}},
  \href{http://www.arXiv.org/abs/1202.4083}{\texttt{ arXiv:1202.4083}}.

\bibitem{Aad:2012yfa}
\hrefCMSnoop {} {{ATLAS Collaboration}, ``{Search for the neutral Higgs bosons
  of the Minimal Supersymmetric Standard Model in pp collisions at $\sqrt{s}=7$
  TeV with the ATLAS detector}'',} (2012).
  \href{http://www.arXiv.org/abs/1211.6956}{\texttt{ arXiv:1211.6956}}.
Submitted to JHEP.
%%CITATION = ARXIV:1211.6956;%%.

\bibitem{Schael:2006cr}
\hrefCMSnoop {} {{ALEPH, DELPHI, L3, and OPAL Collaborations, LEP Working Group
  for Higgs Boson Searches}, S.~Schael, {et~al.}, ``{Search for neutral MSSM
  Higgs bosons at LEP}'',} \textit{ Eur. Phys. J. C} \textbf{ 47} (2006) 547,
  \href{http://dx.doi.org/10.1140/epjc/s2006-02569-7}{\doi{10.1140/epjc/s2006-02569-7}},
\href{http://www.arXiv.org/abs/hep-ex/0602042}{\texttt{ arXiv:hep-ex/0602042}}.
%%CITATION = HEP-EX/0602042;%%.

\bibitem{Aaltonen:2009vf}
\hrefCMSnoop {} {{ CDF} Collaboration, ``{Search for Higgs bosons predicted in
  two-Higgs-doublet models via decays to tau lepton pairs in 1.96-TeV
  $p\overline{p}$ collisions}'',} \textit{ Phys. Rev. Lett.} \textbf{ 103}
  (2009) 201801,
  \href{http://dx.doi.org/10.1103/PhysRevLett.103.201801}{\doi{10.1103/PhysRevLett.103.201801}},
\href{http://www.arXiv.org/abs/0906.1014}{\texttt{ arXiv:0906.1014}}.
%%CITATION = ARXIV:0906.1014;%%.

\bibitem{Abazov:2008hu}
\hrefCMSnoop {} {{ D0} Collaboration, ``{Search for Higgs bosons decaying to
  $\tau$ pairs in $p \bar{p}$ collisions with the D0 detector}'',} \textit{
  Phys. Rev. Lett.} \textbf{ 101} (2008) 071804,
  \href{http://dx.doi.org/10.1103/PhysRevLett.101.071804}{\doi{10.1103/PhysRevLett.101.071804}},
\href{http://www.arXiv.org/abs/0805.2491}{\texttt{ arXiv:0805.2491}}.
%%CITATION = ARXIV:0805.2491;%%.

\bibitem{Abazov2012569}
\hrefCMSnoop {} {{ CMS} Collaboration, ``Search for {H}iggs bosons of the
  minimal supersymmetric standard model $\text{p}\bar{\text{p}}$ in collisions
  at ${\sqrt{s}=1.96\TeV}$'',} \textit{ Phys. Lett. B} \textbf{ 710} (2012)
  569,
  \href{http://dx.doi.org/10.1016/j.physletb.2012.03.021}{\doi{10.1016/j.physletb.2012.03.021}},
\href{http://www.arXiv.org/abs/1112.5431}{\texttt{ arXiv:1112.5431}}.
%%CITATION = ARXIV:1112.5431;%%.

\bibitem{PhysRevD.86.091101}
\hrefCMSnoop {} {{ CDF and D0} Collaboration, ``Search for neutral {H}iggs
  bosons in events with multiple bottom quarks at the {Tevatron}'',} \textit{
  Phys. Rev. D} \textbf{ 86} (2012) 091101,
  \href{http://dx.doi.org/10.1103/PhysRevD.86.091101}{\doi{10.1103/PhysRevD.86.091101}},
  \href{http://www.arXiv.org/abs/1207.2757}{\texttt{ arXiv:1207.2757}}.

\bibitem{Ball:2007zza}
\hrefCMSnoop {} {{ CMS} Collaboration, ``{CMS technical design report, volume
  II: Physics performance}'',} \textit{ J. Phys. G} \textbf{ 34} (2007) 995,
\href{http://dx.doi.org/10.1088/0954-3899/34/6/S01}{\doi{10.1088/0954-3899/34/6/S01}}.
%%CITATION = CERN-LHCC-2006-021 ETC.;%%.

\bibitem{Carena:2011a}
M.~Carena\hrefCMSnoop {} { {et~al.}, ``{LHC} discovery potential for
  non-standard {H}iggs bosons in the 3b channel'',} \textit{ JHEP} \textbf{ 07}
  (2012) 091,
  \href{http://dx.doi.org/10.1007/JHEP07(2012)091}{\doi{10.1007/JHEP07(2012)091}},
  \href{http://www.arXiv.org/abs/1203.1041}{\texttt{ arXiv:1203.1041}}.

\bibitem{Chatrchyan:2008zzk}
\hrefCMSnoop {} {{ CMS} Collaboration, ``The {CMS} experiment at the {CERN}
  {LHC}'',} \textit{ JINST} \textbf{ 3} (2008) S08004,
\href{http://dx.doi.org/10.1088/1748-0221/3/08/S08004}{\doi{10.1088/1748-0221/3/08/S08004}}.
%%CITATION = JINST,3,S08004;%%.

\bibitem{CMS-PAS-PFT-09-001}
\href {http://cdsweb.cern.ch/record/1194487} {{CMS Collaboration},
  ``Particle-flow event reconstruction in {CMS} and performance for jets, taus,
  and {\MET}'',} CMS Physics Analysis Summary CMS-PAS-PFT-09-001, CERN, (2009).

\bibitem{CMS-PAS-PFT-10-001}
\href {http://cdsweb.cern.ch/record/1247373} {{CMS Collaboration},
  ``Commissioning of the particle-flow event reconstruction with the first
  {LHC} collisions recorded in the {CMS} detector'',} CMS Physics Analysis
  Summary CMS-PAS-PFT-10-001, CERN, (2010).

\bibitem{Cacciari:2008gp}
\hrefCMSnoop {} {M.~Cacciari, G.~P. Salam, and G.~Soyez, ``The anti-$k_t$ jet
  clustering algorithm'',} \textit{ JHEP} \textbf{ 4} (2008) 63,
  \href{http://dx.doi.org/10.1088/1126-6708/2008/04/063}{\doi{10.1088/1126-6708/2008/04/063}},
\href{http://www.arXiv.org/abs/0802.1189}{\texttt{ arXiv:0802.1189}}.
%%CITATION = ARXIV:0802.1189;%%.

\bibitem{Chatrchyan:2011ds}
\hrefCMSnoop {} {{ CMS} Collaboration, ``{Determination of jet energy
  calibration and transverse momentum resolution in CMS}'',} \textit{ JINST}
  \textbf{ 6} (2011) P11002,
  \href{http://dx.doi.org/10.1088/1748-0221/6/11/P11002}{\doi{10.1088/1748-0221/6/11/P11002}},
\href{http://www.arXiv.org/abs/1107.4277}{\texttt{ arXiv:1107.4277}}.
%%CITATION = ARXIV:1107.4277;%%.

\bibitem{1748-0221-7-10-P10002}
\hrefCMSnoop {} {{ CMS} Collaboration, ``Performance of {CMS} muon
  reconstruction in pp collision events at {$\sqrt{s} = 7$\TeV}'',} \textit{ J.
  Instrum.} \textbf{ 7} (2012) P10002,
  \href{http://dx.doi.org/10.1088/1748-0221/7/10/P10002}{\doi{10.1088/1748-0221/7/10/P10002}}.

\bibitem{CMS-PAS-BTV-12-001}
\href {http://cdsweb.cern.ch/record/1494669} {{CMS Collaboration},
  ``Identification of b-quark jets with the {CMS} experiment'',} (2012).
  \href{http://www.arXiv.org/abs/1211.4462}{\texttt{ arXiv:1211.4462}}.
  Submitted to \textit{JINST}.

\bibitem{bib:AVF}
\href {http://cdsweb.cern.ch/record/1166320} {W.~Waltenberger, ``Adaptive
  Vertex Reconstruction'',} CMS Note 2008/33, CERN, (2008).

\bibitem{Agostinelli:2002hh}
\hrefCMSnoop {} {{ GEANT4} Collaboration, ``{GEANT4}---a simulation toolkit'',}
  \textit{ Nucl. Instrum. Meth. A} \textbf{ 506} (2003) 250,
  \href{http://dx.doi.org/10.1016/S0168-9002(03)01368-8}{\doi{10.1016/S0168-9002(03)01368-8}}.

\bibitem{Sjostrand:2006za}
\hrefCMSnoop {} {T.~Sj{\"o}strand, S.~Mrenna, and P.~Z. Skands, ``{PYTHIA 6.4
  physics and manual}'',} \textit{ JHEP} \textbf{ 05} (2006) 026,
\href{http://dx.doi.org/10.1088/1126-6708/2006/05/026}{\doi{10.1088/1126-6708/2006/05/026}}.
%%CITATION = HEP-PH/0603175;%%.

\bibitem{Dittmaier:2012vm}
\hrefCMSnoop {} {{ LHC Higgs Cross Section Working Group} Collaboration,
  ``{Handbook of LHC Higgs Cross Sections: 2. Differential Distributions}'',}
  (CERN, Geneva, 2012).
\href{http://www.arXiv.org/abs/1201.3084}{\texttt{ arXiv:1201.3084}}.
%%CITATION = ARXIV:1201.3084;%%.

\bibitem{Mangano:2002ea}
M.~L. Mangano\hrefCMSnoop {} { {et~al.}, ``{ALPGEN, a generator for hard
  multiparton processes in hadronic collisions}'',} \textit{ JHEP} \textbf{ 7}
  (2003) 1,
  \href{http://dx.doi.org/10.1088/1126-6708/2003/07/001}{\doi{10.1088/1126-6708/2003/07/001}},
\href{http://www.arXiv.org/abs/hep-ph/0206293}{\texttt{ arXiv:hep-ph/0206293}}.
%%CITATION = HEP-PH/0206293;%%.

\bibitem{madgraph}
J.~Alwall\hrefCMSnoop {} { {et~al.}, ``MadGraph 5: going beyond'',} \textit{
  JHEP} \textbf{ 06} (2011) 128,
  \href{http://dx.doi.org/10.1007/JHEP06(2011)128}{\doi{10.1007/JHEP06(2011)128}},
  \href{http://www.arXiv.org/abs/1106.0522}{\texttt{ arXiv:1106.0522}}.

\bibitem{CTEQ}
J.~Pumplin\hrefCMSnoop {} { {et~al.}, ``New generation of parton distributions
  with uncertainties from global {QCD} analysis'',} \textit{ JHEP} \textbf{ 7}
  (2002) 12,
  \href{http://dx.doi.org/10.1088/1126-6708/2002/07/012}{\doi{10.1088/1126-6708/2002/07/012}},
  \href{http://www.arXiv.org/abs/hep-ph/0201195}{\texttt{
  arXiv:hep-ph/0201195}}.

\bibitem{Aaltonen:2011nh}
\hrefCMSnoop {} {{ CDF} Collaboration, ``{Search for Higgs Bosons Produced in
  Association with $b$-quarks}'',} \textit{ Phys. Rev. D} \textbf{ 85} (2012)
  032005,
  \href{http://dx.doi.org/10.1103/PhysRevD.85.032005}{\doi{10.1103/PhysRevD.85.032005}},
\href{http://www.arXiv.org/abs/1106.4782}{\texttt{ arXiv:1106.4782}}.
%%CITATION = ARXIV:1106.4782;%%.

\bibitem{CMS-PAS-BTV-11-004}
\href {https://cdsweb.cern.ch/record/1427247} {{CMS Collaboration}, ``b-jet
  identification in the {CMS} experiment'',} CMS Physics Analysis Summary
  CMS-PAS-BTV-11-004, CERN, (2012).

\bibitem{CMS-PAS-BTV-11-001}
\href {http://cdsweb.cern.ch/record/1366061} {{ CMS} Collaboration,
  ``Performance of b-jet identification in {CMS}'',} CMS Physics Analysis
  Summary CMS-PAS-BTV-11-001, CERN, (2011).

\bibitem{ref:pdg}
\hrefCMSnoop {} {{Particle Data Group}, J.~Beringer, {et~al.}, ``Review of
  Particle Physics'',} \textit{ Phys. Rev. D} \textbf{ 86} (2012) 010001,
  \href{http://dx.doi.org/10.1103/PhysRevD.86.010001}{\doi{10.1103/PhysRevD.86.010001}}.

\bibitem{Botje:2011sn}
M.~Botje\hrefCMSnoop {} { {et~al.}, ``The {PDF4LHC Working Group} Interim
  Recommendations'',} (2011).
  \href{http://www.arXiv.org/abs/1101.0538}{\texttt{ arXiv:1101.0538}}.

\bibitem{Cowan:2010js}
G.~Cowan\hrefCMSnoop {} { {et~al.}, ``{Asymptotic formulae for likelihood-based
  tests of new physics}'',} \textit{ Eur. Phys. J. C} \textbf{ 71} (2011) 1554,
  \href{http://dx.doi.org/10.1140/epjc/s10052-011-1554-0}{\doi{10.1140/epjc/s10052-011-1554-0}},
\href{http://www.arXiv.org/abs/1007.1727}{\texttt{ arXiv:1007.1727}}.
%%CITATION = ARXIV:1007.1727;%%.

\bibitem{CLS1}
\hrefCMSnoop {} {A.~L. Read, ``Presentation of search results: the CLs
  technique'',} \textit{ J. Phys. G} \textbf{ 28} (2002) 2693,
  \href{http://dx.doi.org/10.1088/0954-3899/28/10/313}{\doi{10.1088/0954-3899/28/10/313}}.

\bibitem{CLS2}
\hrefCMSnoop {} {T.~Junk, ``Confidence level computation for combining searches
  with small statistics'',} \textit{ Nucl. Instrum. Meth. A} \textbf{ 434}
  (1999) 435,
  \href{http://dx.doi.org/10.1016/S0168-9002(99)00498-2}{\doi{10.1016/S0168-9002(99)00498-2}},
  \href{http://www.arXiv.org/abs/hep-ex/9902006}{\texttt{
  arXiv:hep-ex/9902006}}.

\bibitem{CMS-NOTE-2011-005}
\href {http://cds.cern.ch/record/1379837} {{ATLAS and CMS Collaborations and
  LHC Higgs Combination Group}, ``Procedure for the LHC Higgs boson search
  combination in Summer 2011'',} Technical Report ATL-PHYS-PUB-2011-11,
  CMS-NOTE-2011-005, CERN, (2011).

\bibitem{RooStats}
L.~Moneta\href
  {http://pos.sissa.it/archive/conferences/093/057/ACAT2010_057.pdf} {
  {et~al.}, ``The {R}oo{S}tats {P}roject'',} in \textit{ 13$^\text{th}$
  International Workshop on Advanced Computing and Analysis Techniques in
  Physics Research (ACAT2010)}.
\newblock PoS ACAT:057 (2010).
\newblock \href{http://www.arXiv.org/abs/1009.1003}{\texttt{ arXiv:1009.1003}}.
\newblock {PoS ACAT:057 (2010) }.

\bibitem{ref:mHmax1}
M.~Carena\hrefCMSnoop {} { {et~al.}, ``Suggestions for benchmark scenarios for
  {MSSM} {H}iggs boson searches at hadron colliders'',} \textit{ Eur. Phys. J.
  C} \textbf{ 26} (2003) 601,
  \href{http://dx.doi.org/10.1140/epjc/s2002-01084-3}{\doi{10.1140/epjc/s2002-01084-3}},
  \href{http://www.arXiv.org/abs/hep-ph/0202167}{\texttt{
  arXiv:hep-ph/0202167}}.

\bibitem{ref:mHmax2}
M.~Carena\hrefCMSnoop {} { {et~al.}, ``{MSSM Higgs boson searches at the
  Tevatron and the LHC: Impact of different benchmark scenarios}'',} \textit{
  Eur. Phys. J. C} \textbf{ 45} (2006) 797,
  \href{http://dx.doi.org/10.1140/epjc/s2005-02470-y}{\doi{10.1140/epjc/s2005-02470-y}},
  \href{http://www.arXiv.org/abs/hep-ph/0511023}{\texttt{
  arXiv:hep-ph/0511023}}.

\bibitem{Carena:2013qia}
M.~Carena\hrefCMSnoop {} { {et~al.}, ``{MSSM Higgs Boson Searches at the LHC:
  Benchmark Scenarios after the Discovery of a Higgs-like Particle}'',}
\href{http://www.arXiv.org/abs/hep-ph/1302.7033}{\texttt{
  arXiv:hep-ph/1302.7033}}.
%%CITATION = ARXIV:1302.7033;%%.

\bibitem{Harlander:2003ai}
\hrefCMSnoop {} {R.~V. Harlander and W.~B. Kilgore, ``{Higgs boson production
  in bottom quark fusion at next-to-next-to-leading order}'',} \textit{ Phys.
  Rev. D} \textbf{ 68} (2003) 013001,
  \href{http://dx.doi.org/10.1103/PhysRevD.68.013001}{\doi{10.1103/PhysRevD.68.013001}},
\href{http://www.arXiv.org/abs/hep-ph/0304035}{\texttt{ arXiv:hep-ph/0304035}}.
%%CITATION = HEP-PH/0304035;%%.

\bibitem{Heinemeyer:1998yj}
\hrefCMSnoop {} {S.~Heinemeyer, W.~Hollik, and G.~Weiglein, ``{FeynHiggs: A
  program for the calculation of the masses of the neutral CP even Higgs bosons
  in the MSSM}'',} \textit{ Comput. Phys. Commun.} \textbf{ 124} (2000) 76,
  \href{http://dx.doi.org/10.1016/S0010-4655(99)00364-1}{\doi{10.1016/S0010-4655(99)00364-1}},
\href{http://www.arXiv.org/abs/hep-ph/9812320}{\texttt{ arXiv:hep-ph/9812320}}.
%%CITATION = HEP-PH/9812320;%%.

\bibitem{Heinemeyer:1998np}
\hrefCMSnoop {} {S.~Heinemeyer, W.~Hollik, and G.~Weiglein, ``{The masses of
  the neutral CP-even Higgs bosons in the MSSM: Accurate analysis at the two
  loop level}'',} \textit{ Eur. Phys. J. C} \textbf{ 9} (1999) 343,
  \href{http://dx.doi.org/10.1007/s100529900006}{\doi{10.1007/s100529900006}},
\href{http://www.arXiv.org/abs/hep-ph/9812472}{\texttt{ arXiv:hep-ph/9812472}}.
%%CITATION = HEP-PH/9812472;%%.

\bibitem{Degrassi:2002fi}
G.~Degrassi\hrefCMSnoop {} { {et~al.}, ``{Towards high precision predictions
  for the MSSM Higgs sector}'',} \textit{ Eur. Phys. J. C} \textbf{ 28} (2003)
  133,
  \href{http://dx.doi.org/10.1140/epjc/s2003-01152-2}{\doi{10.1140/epjc/s2003-01152-2}},
\href{http://www.arXiv.org/abs/hep-ph/0212020}{\texttt{ arXiv:hep-ph/0212020}}.
%%CITATION = HEP-PH/0212020;%%.

\bibitem{Frank:2006yh}
M.~Frank\hrefCMSnoop {} { {et~al.}, ``{The Higgs boson masses and mixings of
  the complex MSSM in the Feynman-diagrammatic approach}'',} \textit{ JHEP}
  \textbf{ 02} (2007) 47,
  \href{http://dx.doi.org/10.1088/1126-6708/2007/02/047}{\doi{10.1088/1126-6708/2007/02/047}},
\href{http://www.arXiv.org/abs/hep-ph/0611326}{\texttt{ arXiv:hep-ph/0611326}}.
%%CITATION = HEP-PH/0611326;%%.

\end{thebibliography}\endgroup
